\documentclass[3p,fleqn,a4paper]{elsarticle}

\usepackage{graphicx}
\usepackage{wrapfig}
\usepackage{epsfig}
\usepackage[usenames,dvipsnames]{color}
\usepackage{multirow}
\usepackage{amsfonts,amsmath,amssymb}
\usepackage[hang]{subfigure}
\usepackage{textcomp}
\usepackage{empheq}
\usepackage{algorithm2e}
\RestyleAlgo{ruled}

\usepackage{mathrsfs}
\usepackage{amsthm}
\usepackage{listings}

\usepackage{bbm}
\usepackage{xspace}
\usepackage{yfonts}
\biboptions{sort&compress}
\usepackage{xcolor}

\usepackage[mathlines]{lineno}

\usepackage{caption}
\usepackage{algorithmic}

\usepackage{mathrsfs}
\newcommand{\RN}[1]{%
	\textup{\uppercase\expandafter{\romannumeral#1}}%
}

\usepackage{blindtext}
\journal{---------}

\usepackage{pdfpages}
\usepackage[colorlinks=true,linkcolor=black, citecolor=blue, urlcolor=blue]{hyperref}

\usepackage{natbib}

\begin{document}

\begin{frontmatter}

\title{A hybrid electromechanical phase-field and deep learning framework for predicting fracture in dielectric nanocomposites}

\author[ISD,SUST]{Aamir Dean\corref{cor1}}
\ead{a.dean@isd.uni-hannover.de}

\cortext[cor1]{Corresponding author}

\author[ISD]{Jaykumar Mavani}

\author[ISD]{Betim Bahtiri}

\author[Oslo1,Oslo2]{Behrouz Arash}

\author[ISD]{Raimund Rolfes }

\address[ISD]{Institute of Structural Analysis. Leibniz Universit\"at  Hannover, Appelstr. 9A, 30167 Hannover, Germany}
\address[SUST]{School of Civil Engineering, College of Engineering, Sudan University of Science and Technology, P.O. Box 72, Khartoum, Sudan}
\address[Oslo1]{Department of Mechanical, Electrical, and Chemical Engineering, Oslo Metropolitan University, Pilestredet 35, 0166 Oslo, Norway}
\address[Oslo2]{Green Energy Lab, Department of Mechanical, Electrical and Chemical Engineering, OsloMet - Oslo Metropolitan University, Oslo, Norway}

\begin{abstract}

\textcolor{black}{The accurate and efficient prediction of crack propagation in dielectric materials is a critical challenge in structural health monitoring and the design of smart systems. This work presents a hybrid modeling framework that combines an electromechanical phase-field fracture model with deep learning-based surrogate modeling to predict fracture evolution in dielectric nanocomposite plates. The underlying finite element simulations capture the coupling between mechanical deformation and electrical field perturbations caused by cracks, using a variational phase-field formulation. High-fidelity simulation outputs--namely, phase-field damage variables and electric potential fields--are utilized to train convolutional neural networks (CNNs) with ResNet-U-Net architectures. Crucially, the framework is designed to predict the final crack path directly from the geometric configuration (random defect patterns). We systematically compare the effectiveness of using either phase-field variables or electric potential fields as the primary physical signatures to guide the training process. The results reveal that models informed by electric potential fields offer superior segmentation accuracy, faster convergence, and enhanced generalization, owing to the smoother gradient distribution and global spatial coverage of the electrical response. Ultimately, the trained surrogate model enables the instantaneous, geometry-driven prediction of crack paths, bypassing the need for computationally intensive field calculations during inference. This demonstrates that leveraging electrical physics as a "training guide" significantly improves the reliability of real-time fracture assessment in smart materials.}

\end{abstract}

\end{frontmatter}

\section*{Nomenclature}

\noindent
\textbf{Symbols} \\
\begin{tabular}{p{1cm} p{10cm}}
	$\mathbf{u}$ & Displacement vector\\
	$\boldsymbol{\varepsilon}$ & Infinitesimal strain tensor \\
	$\boldsymbol{\sigma}$ & Cauchy stress tensor\\
	$\mathbf{f}$ & Body force vector\\
	$\phi$ & Electric potential\\
	$\mathbf{E}$ & Electric field vector\\
	$\mathbf{D}$ & Electric displacement vector\\
	$\rho_e$ & Free charge density\\
	$\mathbf{t}$ & Mechanical traction\\
	$q_e$ & Electric surface charge density\\
	$\mathfrak{d}$ & Phase-field damage variable\\
	$\ell$ & Regularization length scale parameter\\
	$G_{c}$ & Critical energy release rate / fracture toughness\\
	$\psi^{\text{mech}}$ & Elastic energy density\\
	$\psi^{\text{elec}}$ & Electrostatic energy density\\
	$\psi^{\text{frac}}$ & Fracture energy density\\
	$\mathbb{C}$ & Fourth-order elasticity tensor\\
	$\lambda, \mu$ & Lamé parameters\\
	$E$ & Young’s modulus\\
	$\nu$ & Poisson’s ratio\\
	$\boldsymbol{\epsilon}$ & Dielectric permittivity tensor\\
	$\epsilon$ & Dielectric permittivity\\
	$\mathcal{H}$ & History field for irreversibility\\
\end{tabular}

\vspace{0.5cm}

\noindent
\textbf{Functions and parameters} \\
\begin{tabular}{p{1cm} p{10cm}}
	$g_1$ & Mechanical degradation function\\
	$g_2$ & Dielectric degradation function\\
	$\kappa_\eta$ & Residual stiffness parameter\\
	$\kappa_\epsilon$ & Residual permittivity parameter\\
	$k, n$ & Degradation parameters\\
\end{tabular}

\vspace{0.5cm}

\noindent
\textbf{Abbreviations} \\
\begin{tabular}{p{1cm} p{10cm}}
	CNN & \textcolor{black}{Convolutional neural network}\\
	U-Net & \textcolor{black}{Encoder–decoder convolutional neural network}\\
	ResNet & Residual neural network\\
	FE & \textcolor{black}{Finite element}\\
	SHM & \textcolor{black}{Structural health monitoring}\\
\end{tabular}

\section{Introduction}
\label{Introduction}

The relentless pursuit of higher performance, efficiency, and safety in modern engineering has led to the widespread adoption of advanced materials, among which polymer composites stand preeminent. Their integration into critical sectors, including the aerospace, automotive, naval, and biomedical industries, is driven by an exceptional combination of high specific stiffness and strength, low density, and superior fatigue endurance when compared to traditional monolithic materials like metal alloys, see \cite{Ref01,Ref02,Ref03}. The demand for these materials continues to grow at an exponential rate, a trend underscored by their expanding utilization in primary aircraft structures and other high-stakes industrial applications \cite{Ref04}. However, the full potential of composites is often tempered by the profound complexity of their failure mechanisms, a direct consequence of their inherent heterogeneity and anisotropy \cite{Ref046}.

The paradigm of composite materials is currently undergoing another significant evolution with the advent of multifunctional or "smart" composites, which are engineered to perform tasks beyond their primary structural role, such as sensing, actuation, or self-healing, see \cite{Ref05,Ref06,Ref07} among others. A particularly promising class of these materials for structural health monitoring (SHM) applications is the dielectric nanocomposite. These are typically formed by embedding dielectric nanofillers, such as copper phthalocyanine (CuPc), into a polymer matrix to create a dielectric elastomer--a type of electroactive polymer that exhibits strong electromechanical coupling \cite{Ref08}. This coupling manifests in two ways: the material deforms when subjected to an electric field, and conversely, its electrical properties--notably its dielectric permittivity and capacitance--change in response to mechanical strain or damage \cite{Ref09}. This latter effect forms the basis of their application in SHM. By continuously monitoring the electrical response of the material, it becomes possible to detect, locate, and even quantify the extent of damage in real-time, creating a "self-sensing" structure that can diagnose its own health condition \cite{Ref07,Ref010}. This capability obviates the need for many traditional, often cumbersome and offline, non-destructive testing (NDT) techniques and is paramount for ensuring the operational safety and reliability of structures where failure could be catastrophic. This integration of sensing functionality directly into the load-bearing material introduces a fundamental duality that poses a profound modeling challenge. In these dielectric nanocomposites, the material is simultaneously the structure and the sensor  \cite{Ref010}. A fracture event, such as the initiation and propagation of a crack, is therefore not merely a mechanical failure; it is intrinsically a coupled electromechanical phenomenon. The crack degrades the mechanical integrity of the composite by reducing its stiffness and strength, while at the same time, it perturbs the local electrical field by altering the dielectric constant in the damaged region, thereby directly impacting the sensing functionality, see  \cite{Ref012,Ref011}. Consequently, any predictive model that aims for physical accuracy cannot treat these aspects in isolation. A simple mechanical fracture model is insufficient because it neglects the electrical response, and a simple electrical model is insufficient because it ignores the mechanical failure driving the change. A fully coupled electromechanical fracture model is therefore not just an academic exercise but a fundamental necessity for understanding the behavior, predicting the reliability, and guiding the design of these advanced multifunctional material systems.

Given the intricate nature of damage evolution in composites, purely experimental characterization is often insufficient for achieving reliable design and life prediction  \cite{Ref013}. This has established computational modeling as an indispensable tool in the field of fracture mechanics. Over the past two decades, the phase-field (PF) method  has emerged as a particularly powerful and versatile computational framework, overcoming many of the significant algorithmic and implementation challenges associated with classical discrete fracture approaches, such as the extended finite element method (XFEM), see \cite{Ref014,Ref015,Ref016,Ref038,Ref039,Ref040,Ref041,Ref042,Ref053,Ref054} among others. The core concept of PF method is to abandon the explicit representation of cracks as sharp geometric discontinuities. Instead, a crack is represented in a diffuse or "smeared" manner over a narrow region by introducing a continuous scalar variable, known as the phase field, typically denoted by '$\mathfrak{d}$'. This field acts as an order parameter, smoothly transitioning from an undamaged state (e.g., '$\mathfrak{d}=0$') to a fully fractured state ('$\mathfrak{d}=1$'), see \cite{Ref017,Ref018,Ref025} and the references cited therein. This diffusive representation is governed by a partial differential equation derived from a variational energy principle. This allows the model to naturally and robustly predict complex fracture phenomena, including the initiation of cracks from arbitrary sites, their subsequent propagation along complex paths, and events like crack branching and coalescence, all without the need for ad-hoc criteria or complex topological tracking and remeshing algorithms.

The theoretical underpinnings of PF method are robust, originating from the variational theory of brittle fracture proposed by Francfort and Marigo, which recasts fracture as a global energy minimization problem \cite{Ref017}. The phase-field formulation can be rigorously shown to be a regularized approximation of this foundational theory \cite{Ref018}. The great flexibility of the PF method framework, where the system's behavior is governed by an energy functional, allows for its extension to coupled multi-physics problems by incorporating additional energy terms, see \cite{Ref011,Ref015,Ref019}. This capability has been leveraged to develop phase-field models for fracture in piezoelectric and ferroelectric materials, which, like the dielectric nanocomposites of interest here, exhibit strong electromechanical coupling  \cite{Ref012,Ref011}. These advanced models couple the variational formulation of fracture with the governing equations of electrostatics. A key challenge in this domain is the accurate modeling of the electrical boundary conditions on the crack faces, as the behavior depends on the electrical properties of the medium (e.g., air, vacuum) filling the crack gap. Different physical scenarios must be considered, such as an electrically impermeable (insulating) crack, a permeable crack, or a semi-permeable crack  \cite{Ref020}. In the PF method framework, these conditions are encoded by selectively degrading different terms within the total electromechanical enthalpy density function based on the value of the phase-field variable  \cite{Ref012,Ref011}. A physically consistent model must therefore capture the simultaneous degradation of both the mechanical stiffness and the dielectric permittivity as the crack propagates, introducing significant nonlinearity and mathematical complexity into the governing equations.

While the phase-field method provides the necessary physical fidelity to capture these complex, coupled phenomena, this power comes at a steep and often prohibitive price: extreme computational cost \cite{Ref021}. This "curse of fidelity" arises from several compounding factors. First, a standard PF method simulation already necessitates the use of a very fine computational mesh to accurately resolve the narrow, diffuse crack region, as the mesh size must be smaller than the phase-field length scale parameter. Second, when coupling with another physical field like electrostatics, a new set of governing equations and variables (e.g., electric potential) must be solved at every node in the mesh, increasing the size of the algebraic system. Third, and most critically, the coupling terms in the energy functional introduce strong nonlinearities, requiring more iterations within each time step for the numerical solver to converge. The combination of these factors leads to a multiplicative, rather than additive, increase in computational expense. As a result, high-fidelity, coupled electromechanical phase-field simulations can be so computationally intensive that they become impractical for tasks such as large-scale structural analysis, iterative design optimization, or real-time damage assessment, which are precisely the applications for which smart materials are intended. This establishes a critical research gap: a powerful theoretical tool exists to model the physics correctly, but it is too slow to be of practical use. There is a pressing need for a new computational paradigm that can retain the physical accuracy of coupled PF method while drastically reducing the associated computational burden.

To surmount the computational bottleneck imposed by high-fidelity simulations, the scientific community is increasingly turning to machine learning (ML) and deep learning (DL). The central strategy is to develop a data-driven "surrogate" model--a trained neural network that learns to approximate the complex -output relationship of the expensive physics-based simulation but can make predictions at a fraction of the computational cost \cite{Ref023,Ref024,Ref027,Ref028,Ref029,Ref049}. The application of ML, particularly artificial neural networks (ANNs), to problems in fracture mechanics is a rapidly expanding field, see \cite{Ref023,Ref024,Ref026,Ref031} among others. Early efforts have demonstrated the ability of ANNs to predict key fracture parameters, such as the stress intensity factor (SIF) for cracked plates, by training the networks on large datasets generated via conventional finite element analysis (FEA).

For problems in continuum mechanics, where the outputs are spatially distributed fields (e.g., stress, strain, displacement, or phase-field), convolutional neural networks (CNNs) have emerged as a particularly suitable architecture. Originally developed for and famously successful in the domain of image recognition, CNNs are inherently designed to process grid-like data and learn hierarchical spatial features, see \cite{Ref030,Ref032,Ref033} and the references cited therein. The convolutional layers of a CNN act as learnable filters that slide across the input data, identifying local patterns. This mechanism is conceptually analogous to how physical phenomena, such as stress concentrations, manifest locally within a material domain, for instance, around the tip of a crack \cite{Ref034}. The efficacy of this approach has been demonstrated in several studies where CNNs have been successfully employed as surrogate models to predict the effective mechanical properties of periodic composites, see \cite{Ref035,Ref036,Ref037} among many others. In these works, the material microstructure is represented as a binary image, and the CNN is trained to regress the homogenized properties, achieving prediction speeds that are orders of magnitude faster than direct FEA simulations.

\textcolor{black}{Recently, a growing body of research has focused on hybrid frameworks that combine phase-field  fracture models with machine learning or deep learning techniques to accelerate fracture simulations or to infer damage evolution from high-fidelity data. In these studies, deep neural networks have been trained as surrogate solvers to approximate the solution of the phase-field governing equations, significantly reducing computational cost while retaining acceptable accuracy. Representative works include the use of convolutional neural networks to predict phase-field evolution in brittle fracture problems, recurrent or encoder–decoder architectures for time-dependent crack growth, and physics-informed or data-driven approaches to identify fracture parameters and crack paths in purely mechanical settings, see \cite{Ref023,Ref024,Ref026} among many others. These contributions have demonstrated the promise of PF–DL coupling for fast fracture prediction, particularly in elastic and quasi-brittle materials under mechanical loading.}

\textcolor{black}{In contrast to existing works, the present study advances the state of the art in two fundamental directions. First, we develop a hybrid deep learning framework built upon a fully coupled electromechanical phase-field formulation for dielectric nanocomposites, enabling the surrogate model to learn fracture behavior in a genuinely multi-physics setting. This allows the network to implicitly capture the simultaneous mechanical degradation and dielectric permittivity evolution induced by crack growth--a feature essential for modeling smart, self-sensing composite materials. Second, we introduce the electric potential field as a novel physical signature to guide the training of crack segmentation models. Rather than training the network solely on local strain or phase-field variables, we demonstrate that leveraging the electric potential--as a global and smoothly varying physical response--provides richer gradient information during the learning process. This leads to superior convergence, accuracy, and generalization, even when the final model performs predictions based solely on geometric configurations. To the authors’ knowledge, this is the first study to systematically exploit electric potential fields as the primary physical signal to inform geometry-to-crack mapping within a PF-DL framework.}

The accurate and efficient prediction of fracture in multifunctional dielectric nanocomposites is essential for their reliable design and deployment in next-generation technologies, particularly for real-time structural health monitoring. While coupled electromechanical phase-field models possess the requisite physical fidelity to capture the complex failure behavior of these materials, their prohibitive computational expense renders them impractical for the very applications they are meant to enable. This creates a critical disconnect between modeling capability and practical utility.
To bridge this gap, this study proposes and validates a novel hybrid computational framework that synergistically combines high-fidelity, fully coupled electromechanical phase-field fracture modeling with deep learning-based surrogate modeling. By training a deep neural network on data generated from the physics-based PF method, the framework aims to achieve the predictive accuracy of the high-fidelity model at a computational speed amenable to large-scale analysis and design. The primary objectives and specific contributions of this paper are fourfold: (i) to develop and implement a thermodynamically consistent electromechanical phase-field fracture model for dielectric nanocomposites, capable of capturing the coupled degradation of mechanical stiffness and dielectric permittivity under electromechanical loading, (ii) to utilize this high-fidelity simulation framework to generate a comprehensive, high-resolution dataset of simulation results. This data set, which encompasses the evolution of phase-field and electric potential variables, serves as the ground-truth data for training the deep learning surrogate model, (iii) \textcolor{black}{to design and train a deep CNN surrogate based on ResNet-U-Net architectures that learns to map initial geometric configurations (random defect patterns) directly to final crack paths. This stage leverages two distinct training strategies--one informed by phase-field data and the other by electric potential fields arising from electromechanical coupling--to assess which physical signature better guides the network in learning the underlying fracture mechanics}, and (iv) to conduct a rigorous evaluation of the surrogate model’s performance, assessing its predictive accuracy on unseen scenarios and generalization capability.

\section{Theoretical background and methodology}
\label{Theoretical}

\subsection{Electromechanical phase-field fracture model}
\label{Electromechanical}

This section presents a coupled electromechanical phase-field fracture formulation under small-strain assumptions, describing the degradation of mechanical and electrical fields due to crack nucleation and evolution in electro-active materials. \textcolor{black}{The governing equations, constitutive relations, energy functional, and weak form are formulated within a variational framework for coupled electromechanical fracture. Detailed derivations and numerical implementation are referred to \cite{Ref011}.}

\subsubsection{Kinematics and balance laws}
\label{Kinematics}

Let $\Omega \subset \mathbb{R}^3$ denote a bounded domain representing the reference configuration of the electro-mechanical body, with boundary $\partial \Omega$. The displacement field is $\mathbf{u} : \Omega \rightarrow \mathbb{R}^3$, and the electric potential is $\phi : \Omega \rightarrow \mathbb{R}$. The body is subjected to a mechanical traction $\mathbf{t}$ and an electric surface charge density $q_e$ on the boundary.

Under small strain assumptions, the infinitesimal strain tensor $\boldsymbol{\varepsilon}$ is defined as:
\begin{equation}
    \boldsymbol{\varepsilon}(\mathbf{u}) = \frac{1}{2} \left( \nabla \mathbf{u} + (\nabla \mathbf{u})^T \right).
    \label{Eq1}
\end{equation}

The electric field $\mathbf{E}$ is derived from the gradient of the electric potential:
\begin{equation}
    \mathbf{E} = -\nabla \phi.
    \label{Eq2}
\end{equation}

The governing equations for mechanical equilibrium and electrostatics (Gauss's law in differential form) are given respectively by:
\begin{equation}
    \nabla \cdot \boldsymbol{\sigma} + \mathbf{f} = \mathbf{0} \quad \text{in } \Omega,
    \label{Eq3}
\end{equation}
\begin{equation}
    \nabla \cdot \mathbf{D} = \rho_e \quad \text{in } \Omega,
    \label{Eq4}
\end{equation}

where $\boldsymbol{\sigma}$ is the Cauchy stress tensor, $\mathbf{f}$ is the body force vector, $\mathbf{D}$ is the electric displacement vector, and $\rho_e$ is the free charge density.

\subsubsection{Energy-based phase-field formulation}
\label{Energy}

\textcolor{black}{Crack initiation and propagation in electro-active materials are modeled using a phase-field approach, in which the sharp crack topology $\Gamma$ is regularized by the scalar fracture variable $\mathfrak{d}$ and coupled to the mechanical and electrical fields. The total free energy accounts for the degradation of both fields as well as the fracture energy.}

The total free energy functional $\mathcal{E}$ of the coupled system is postulated as:
\begin{equation}
    \mathcal{E}[\mathbf{u}, \phi, \mathfrak{d}] = \int_{\Omega} \left( \psi^{\text{mech}}(\boldsymbol{\varepsilon}, \mathfrak{d}) + \psi^{\text{elec}}(\mathbf{E}, \mathfrak{d}) + \psi^{\text{frac}}(\mathfrak{d}, \nabla \mathfrak{d}) \right) \, \mathrm{d}\Omega - \Pi_{\text{ext}},
    \label{Eq5}
\end{equation}

where $\mathfrak{d}:\Omega \rightarrow [0,1]$ is the scalar phase-field variable describing the fracture state ($\mathfrak{d} = 0$ for intact, $\mathfrak{d} = 1$ for fully broken), $\psi^{\text{mech}}$ is the degraded elastic energy density, $\psi^{\text{elec}}$ is the degraded electrostatic energy density, $\psi^{\text{frac}}$ is the regularized fracture energy density, and $\Pi_{\text{ext}}$ is the potential of external work.

To model fracture effects, both the mechanical and dielectric responses are degraded as a function of the phase-field variable. Commonly used degradation functions for the mechanical and dielectric responses are respectively given by:
\begin{equation}
    g_1(\mathfrak{d}) = (1 - \mathfrak{d})^2 + \kappa_\eta,
    \label{Eq6}
\end{equation}
\begin{equation}
    g_{2}(\mathfrak{d},k,n)=\frac{1-\exp\left(-k\left(1-\mathfrak{d}\right)^{n}\right)}{1-\exp\left(-k\right)} + \kappa_\epsilon,
    \label{Eq6b}
\end{equation}

where $\kappa_\eta$ and $\kappa_\epsilon$ are residual  parameters to prevent ill-conditioning, $k$ and $n$ are parameters that determine how the degradation function behaves. \textcolor{black}{The exponential dielectric degradation function $g_2(\mathfrak{d})$ is adopted from \cite{Ref011}, where it was shown to effectively capture the nonlinear reduction of permittivity observed in electroactive and dielectric solids under damage. Unlike commonly used quadratic forms, this two-parameter formulation provides additional flexibility through $k$ and $n$ to control the onset and rate of dielectric degradation independently of the mechanical response. The function satisfies the essential physical requirements, ensuring boundedness and a smooth transition between intact and fully fractured states.}

\textcolor{black}{The crack in this framework is represented by the phase-field variable $\mathfrak{d} \in [0,1]$, which degrades both the elastic stiffness and the dielectric permittivity through the degradation function $\boldsymbol\epsilon(\mathfrak{d}) = g_2(\mathfrak{d})\boldsymbol\epsilon_0$, where $g_2(1) = \kappa_\epsilon \ll 1$. Because the degraded permittivity satisfies $\kappa_\epsilon > 0$, the electric field inside fully damaged regions ($\mathfrak{d} \approx 1$) does not vanish entirely; instead, a small residual electric field may penetrate the crack volume. Physically, this corresponds to a crack that is not a perfect vacuum gap but rather a partially bridged or micro-cracked zone, which is common in nanocomposites due to filler ligaments, see \cite{Ref050,Ref052}.}

For isotropic linear elastic materials, the degraded elastic energy density is:
\begin{equation}
    \psi^{\text{mech}}(\boldsymbol{\varepsilon}, \mathfrak{d}) = g_1(\mathfrak{d}) \cdot \frac{1}{2} \boldsymbol{\varepsilon} : \mathbb{C} : \boldsymbol{\varepsilon},
    \label{Eq7}
\end{equation}

where $\mathbb{C}$ is the fourth-order elasticity tensor defined by the Lamé parameters $\lambda$ and $\mu$, or equivalently by Young’s modulus $E$ and Poisson’s ratio $\nu$.

The degraded electric energy density is given by:
\begin{equation}
    \psi^{\text{elec}}(\mathbf{E}, \mathfrak{d}) = g_2(\mathfrak{d}) \cdot \frac{1}{2} \mathbf{E} \cdot \boldsymbol{\epsilon} \cdot \mathbf{E},
    \label{Eq8}
\end{equation}
where $\boldsymbol{\epsilon}$ is the permittivity tensor of the material (typically isotropic: $\boldsymbol{\epsilon} = \epsilon \mathbf{I}$, where $\epsilon$ denotes the dielectric permittivity and $\mathbf{I}$ is the second-order identity tensor).

The regularized fracture energy density based on the Ambrosio–Tortorelli functional is:
\begin{equation}
    \psi^{\text{frac}}(\mathfrak{d},\nabla\mathfrak{d})=G_{c}\left(\frac{\mathfrak{d}^{2}}{2\ell}+\frac{\ell}{2}|\nabla\mathfrak{d}|^{2}\right),
    \label{Eq8b}
\end{equation}

where $G_c$ is the critical energy release rate (fracture toughness) and $\ell$ is the regularization length scale controlling the width of the diffusive crack zone.

\subsubsection{Variational principles and Euler–Lagrange equations}
\label{Variational}

The total potential energy is minimized with respect to the displacement $\mathbf{u}$, electric potential $\phi$, and phase-field $\mathfrak{d}$:
\begin{equation}
    \delta \mathcal{E} = 0 \quad \forall \, \delta \mathbf{u}, \delta \phi, \delta \mathfrak{d}.
    \label{Eq9}
\end{equation}

This yields the coupled system of Euler–Lagrange equations, for mechanical equilibrium, electrostatic equilibrium, and phase-field evolution, respectively:

\begin{equation}
    \nabla \cdot (g_1(\mathfrak{d}) \, \boldsymbol{\sigma}_0) + \mathbf{f} = \mathbf{0},
    \label{Eq10}
\end{equation}

\begin{equation}
    \nabla \cdot (g_2(\mathfrak{d}) \, \mathbf{D}_0) = \rho_e,
    \label{Eq11}
\end{equation}

\begin{equation}
    G_{c}\left(\ell\nabla^{2}\mathfrak{d}-\frac{\mathfrak{d}}{\ell}\right)-g_{1}'(\mathfrak{d})\psi_{0}^{\text{mech}} = 0,
    \label{Eq12}
\end{equation}

where $\psi_{0}^{\text{mech}}=\frac{1}{2}\boldsymbol{\varepsilon}:\mathbb{C}:\boldsymbol{\varepsilon}$.

Optionally, an irreversibility condition can be enforced via a history field $\mathcal{H}(\mathbf{x}, t)$, ensuring that the crack cannot heal:
\begin{equation}
   \mathfrak{d}_t \geq 0 \quad \text{or} \quad \psi^{\text{mech}}_0 \rightarrow \mathcal{H} \quad \text{with} \quad \mathcal{H}_t \geq 0.
   \label{Eq13}
\end{equation}

The system is supplemented with appropriate boundary conditions:
\begin{equation}
   \mathbf{u} = \bar{\mathbf{u}} \text{ on } \partial \Omega_u, \quad (g_1(\mathfrak{d}) \, \boldsymbol{\sigma}_0) \cdot \mathbf{n} = \bar{\mathbf{t}} \text{ on } \partial \Omega_t,
   \label{Eq14}
\end{equation}
\begin{equation}
   \phi = \bar{\phi} \text{ on } \partial \Omega_{\phi}, \quad (g_2(\mathfrak{d}) \, \mathbf{D}_0) \cdot \mathbf{n} = \bar{q}_e \text{ on } \partial \Omega_q ,
   \label{Eq15}
\end{equation}
\begin{equation}
      \nabla \mathfrak{d} \cdot \mathbf{n} = 0 \text{ on } \partial \Omega.
   \label{Eq16}
\end{equation}

\textcolor{black}{The interactions between the mechanical, electrical, and damage fields are summarized in Table~\ref{tab:physics_summary}. Physically, the elastic strain energy density acts as the primary driving force for the evolution of the phase-field. As damage accumulates, the local degradation of the material stiffness and dielectric permittivity redistributes the mechanical stresses and alters the electric potential distribution.}

\begin{table}[h!]
	\centering
	\captionsetup{justification=centering} 
	\caption{\textcolor{black}{Summary of field variables, governing equations, and coupling mechanisms in the electromechanical phase-field framework.}}
	\label{tab:physics_summary}
	\begin{tabular}{llll}
		\hline
		Physical field & Variable & Governing equation & Boundary conditions \\ 
		\hline
		\hline
		Mechanical & $\mathbf{u}$ & Momentum balance (Eq.~\ref{Eq10}) & Prescribed $\bar{\mathbf{u}}$ or $\bar{\mathbf{t}}$ \\
		Electrical & $\phi$ & Gauss's law (Eq.~\ref{Eq11}) & Prescribed $\bar{\phi}$ or $\bar{q}_e$ \\ 
		Damage & $\mathfrak{d}$ & Phase-field evolution (Eq.~\ref{Eq12}) & No-flux ($\nabla \mathfrak{d} \cdot \mathbf{n} = 0$) \\
		\hline
		Coupling & \multicolumn{3}{l}{Physical Mechanism} \\ 
		\hline
		\hline
		Mech. $\to$ Damage & \multicolumn{3}{l}{Elastic energy $\psi_0^{\text{mech}}$ acts as the driving force for $\mathfrak{d}$ evolution.} \\
		Damage $\to$ Mech. & \multicolumn{3}{l}{Stiffness tensor $\mathbb{C}$ is degraded by $g_1(\mathfrak{d})$, altering the stress state.} \\
		Damage $\to$ Elec. & \multicolumn{3}{l}{Permittivity $\boldsymbol{\epsilon}$ is degraded by $g_2(\mathfrak{d})$, imposing insulating conditions.} \\ 
		\hline
	\end{tabular}
\end{table}

\subsubsection{Weak formulation}
\label{Weak}

Let $\Omega \subset \mathbb{R}^3$ denote the reference configuration of the domain, and let $\partial \Omega = \partial \Omega_u \cup \partial \Omega_t = \partial \Omega_\phi \cup \partial \Omega_q$ denote the decomposition of the boundary with respect to essential and natural conditions for displacement and electric potential, respectively.

Let us  consider again the total potential energy functional:
\begin{equation}
    \mathcal{E}[\mathbf{u}, \phi, \mathfrak{d}] = \int_{\Omega} \left( \psi^{\text{mech}}(\boldsymbol{\varepsilon}, \mathfrak{d}) + \psi^{\text{elec}}(\mathbf{E}, \mathfrak{d}) + \psi^{\text{frac}}(\mathfrak{d}, \nabla \mathfrak{d}) \right) \, \mathrm{d}\Omega - \Pi_{\text{ext}}.
    \label{Eq17}
\end{equation}

We define the following admissible function spaces for the displacement field, electric potential, phase-field, respectively: 
\begin{equation}
    \mathbf{u} \in \mathcal{V}_u := \{ \mathbf{v} \in [H^1(\Omega)]^3 \mid \mathbf{v} = \bar{\mathbf{u}} \text{ on } \partial \Omega_u \},
    \label{Eq18}
\end{equation}
\begin{equation}
    \phi \in \mathcal{V}_\phi := \{ \psi \in H^1(\Omega) \mid \psi = \bar{\phi} \text{ on } \partial \Omega_\phi \},
    \label{Eq19}
\end{equation}
\begin{equation}
    \mathfrak{d} \in \mathcal{V}_\mathfrak{d} := \{ w \in H^1(\Omega) \mid 0 \leq w \leq 1 \, \text{a.e. in } \Omega \},
    \label{Eq20}
\end{equation}

We obtain the weak form by taking the first variation of $\mathcal{E}[\mathbf{u}, \phi, \mathfrak{d}]$ with respect to the three fields. For the mechanical equilibrium (displacement field), let us start with:
\begin{equation}
    \delta_{\mathbf{u}} \mathcal{E} = \int_{\Omega} g_1(\mathfrak{d}) \, \boldsymbol{\sigma}_0 : \delta \boldsymbol{\varepsilon} \, \mathrm{d}\Omega - \int_{\Omega} \mathbf{f} \cdot \delta \mathbf{u} \, \mathrm{d}\Omega - \int_{\partial \Omega_t} \bar{\mathbf{t}} \cdot \delta \mathbf{u} \, \mathrm{d}\partial\Omega.
    \label{Eq21}
\end{equation}

Using:
\begin{equation}
    \delta \boldsymbol{\varepsilon} = \frac{1}{2} \left( \nabla \delta \mathbf{u} + (\nabla \delta \mathbf{u})^T \right),
    \label{Eq22}
\end{equation}

and symmetry of $\boldsymbol{\sigma}$, we write:
\begin{equation}
    \int_{\Omega} g_1(\mathfrak{d}) \, \boldsymbol{\sigma}_0 : \delta \boldsymbol{\varepsilon} \, \mathrm{d}\Omega  = \int_{\Omega} g_1(\mathfrak{d}) \, \boldsymbol{\sigma}_0 : \nabla^s \delta \mathbf{u} \, \mathrm{d}\Omega.
    \label{Eq23}
\end{equation}

Thus, the weak form of the mechanical equilibrium equation is:
\begin{equation}
    \boxed{ \int_{\Omega} g_1(\mathfrak{d}) \, \boldsymbol{\sigma}_0 : \nabla^s \delta \mathbf{u} \, \mathrm{d}\Omega = \int_{\Omega} \mathbf{f} \cdot \delta \mathbf{u} \, \mathrm{d}\Omega + \int_{\partial \Omega_t} \bar{\mathbf{t}} \cdot \delta \mathbf{u} \, \mathrm{d}\partial\Omega \quad \forall \, \delta \mathbf{u} \in \mathcal{W}_u}
    \label{Eq24}
\end{equation}

For the electrostatic equilibrium (electric potential field), the variation with respect to the electric potential yields:
\begin{equation}
    \delta_{\phi} \mathcal{E} = \int_{\Omega} g_2(\mathfrak{d}) \, \mathbf{D}_0 \cdot \delta \mathbf{E} \, \mathrm{d}\Omega - \int_{\Omega} \rho_e \, \delta \phi \, \mathrm{d}\Omega  - \int_{\partial \Omega_q} \bar{q}_e \, \delta \phi \, \mathrm{d}\partial\Omega,
    \label{Eq25}
\end{equation}

with:
\begin{equation}
    \delta \mathbf{E} = -\nabla \delta \phi \quad \text{and} \quad \mathbf{D}_0 = {\epsilon} \mathbf{E},
    \label{Eq26}
\end{equation}

we get:
\begin{equation}
    \int_{\Omega} g_2(\mathfrak{d}) \, \boldsymbol{\epsilon} \mathbf{E} \cdot (-\nabla \delta \phi) \, \mathrm{d}\Omega  = -\int_{\Omega} g_2(\mathfrak{d}) \, \epsilon \nabla \phi \cdot \nabla \delta \phi \, \mathrm{d}\Omega.
    \label{Eq27}
\end{equation}

Thus, the weak form of the electrostatic equilibrium equation becomes:
\begin{equation}
    \boxed{ \int_{\Omega} g_2(\mathfrak{d}) \, {\epsilon} \nabla \phi \cdot \nabla \delta \phi \, \mathrm{d}\Omega = \int_{\Omega} \rho_e \, \delta \phi \, \mathrm{d}\Omega + \int_{\partial \Omega_q} \bar{q}_e \, \delta \phi \, \mathrm{d}\partial\Omega \quad \forall \, \delta \phi \in \mathcal{W}_\phi } 
    \label{Eq28}
\end{equation}

Finally, for the phase-field evolution (fracture field), the variation of the total energy with respect to the phase-field $\mathfrak{d}$ yields:
\begin{equation}
    \delta_\mathfrak{d} \mathcal{E} = \int_{\Omega} \left[ g_1'(\mathfrak{d}) \psi^{\text{mech}}_0 \delta d \textcolor{black}{+ g_2'(\mathfrak{d}) \psi^{\text{elec}}_0 \delta \mathfrak{d}} + G_{c}\left(\frac{\mathfrak{d}}{\ell}\delta\mathfrak{d}+\ell\nabla\mathfrak{d}\cdot\nabla\delta\mathfrak{d}\right) \right] \, \mathrm{d}\Omega,
    \label{Eq29}
\end{equation}

\textcolor{black}{where $\psi_{0}^{\text{elec}}=\frac{1}{2} \mathbf{E} \cdot \boldsymbol{\epsilon} \cdot \mathbf{E}$.}

This gives the weak form of the phase-field evolution equation:
\begin{equation}
    \boxed{ \int_{\Omega} \left[ g_1'(\mathfrak{d}) \psi^{\text{mech}}_0 \delta \mathfrak{d} \textcolor{black}{+ g_2'(\mathfrak{d}) \psi^{\text{elec}}_0 \delta \mathfrak{d}}  + G_{c}\left(\frac{\mathfrak{d}}{\ell}\delta\mathfrak{d}+\ell\nabla\mathfrak{d}\cdot\nabla\delta\mathfrak{d}\right) \right] \, \mathrm{d}\Omega = 0 \quad \forall \, \delta d \in \mathcal{W}_d } 
    \label{Eq30}
\end{equation}

The weak forms above are discretized using the finite element method (FEM), and the resulting system of equations is implemented in the general-purpose FEM software \texttt{ABAQUS} using a monolithic solution strategy. For the sake of brevity, the details of FEM discretization and implementation are omitted here. For further details, the reader is referred to \cite{Ref011,Ref019}.

\subsection{Machine learning framework for crack segmentation}
\label{sec:ml_framework}

To create an efficient surrogate model capable of predicting complex fracture paths, we developed a deep learning framework based on a specialized convolutional neural network. The chosen architecture is a U-Net, which is particularly well-suited for image-to-image translation tasks such as semantic segmentation, where the output is a pixel-wise classification map of the input. This section provides a detailed description of the network's architecture, its components, and the underlying principles that make it effective for this application.

\subsubsection{Architectural overview: The U-Net model}
\label{Architectural}
\begin{figure}
\begin{center}
\includegraphics[width=0.8\linewidth]{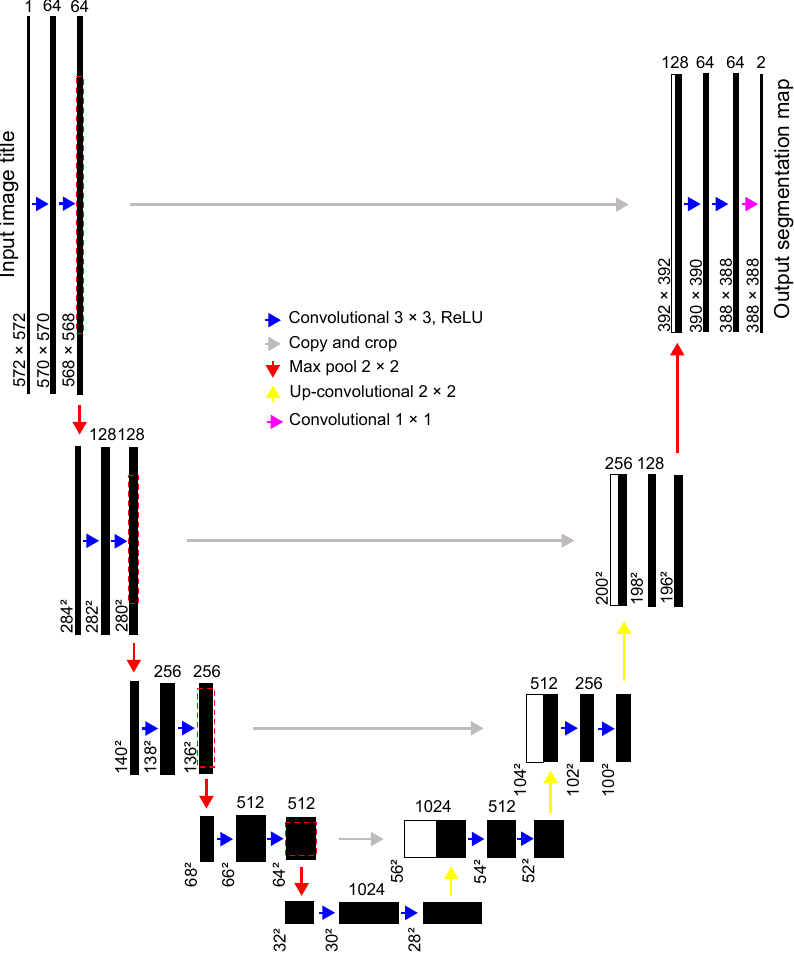}
\caption{Schematic overview of the U-Net architecture employed for crack segmentation, adapted from \cite{Ref043}.}
\label{figure0}
\end{center}
\end{figure}

The U-Net architecture, first proposed for biomedical image segmentation by Ronneberger et al. \cite{Ref043}, is characterized by its symmetric encoder-decoder structure, often visualized in a ‘U’ shape. This design excels at capturing both high-level contextual information and fine-grained spatial details, which are essential for precisely localizing features like crack boundaries. The full architecture is illustrated in Figure~\ref{figure0}.

The input to the network is a single-channel grayscale image of size $572 \times 572$ pixels. The left side of the U is the encoder (or contracting path), which functions as a feature extractor. It consists of repeated blocks of two $3 \times 3$ convolutional layers (with no padding), each followed by a rectified linear unit (ReLU) activation function. After each block, a $2 \times 2$ max pooling operation with a stride of 2 is applied, halving the spatial dimensions while doubling the number of feature channels. This progressively increases the receptive field, allowing the network to build a hierarchical understanding of the image.
 
The right side of the U is the decoder (or expansive path), which reconstructs the segmentation map from the compressed features. Each stage begins with a $2 \times 2$ transposed convolution (up-convolution) that doubles the spatial resolution and halves the number of channels, followed by concatenation with the corresponding feature map from the encoder path. This is followed by two $3 \times 3$ convolutional layers with ReLU activation.

At the bottom of the U is the bottleneck, which marks the transition between encoding and decoding. It consists of a convolutional block that processes the most abstracted representation of the input. The final layer applies a $1 \times 1$ convolution to reduce the number of output channels to two, producing a $388 \times 388$ segmentation map. A softmax activation is applied to obtain pixel-wise classification probabilities.

\subsubsection{Feature extraction enhancement via ResNet backbone}
\label{Feature_extraction }
Instead of a standard "plain" CNN encoder, our U-Net implementation integrates pre-trained residual networks (ResNet) as the encoder backbone. This strategic choice offers two significant advantages. First, it leverages transfer learning, initializing the encoder with weights learned from the vast ImageNet dataset. This provides the network with a powerful, pre-existing knowledge base of generic visual features (e.g., edges, textures, shapes), which accelerates convergence and often leads to better final performance.

Second, the ResNet architecture itself is fundamentally more powerful for training very deep networks. Its core innovation is the "residual block," where the output of a layer (or a stack of layers) is added to its original input via a "skip" or "identity" connection. This allows the network to learn residual functions, effectively letting gradients flow more easily through the deep network during backpropagation. This structure elegantly solves the vanishing gradient problem that plagues traditional deep CNNs, enabling the construction of much deeper and more expressive models. \textcolor{black}{We systematically evaluated four ResNet configurations: ResNet34, ResNet50, ResNet101, and ResNet152. In this notation, the numbers (e.g., 34, 152) correspond to the total depth
of the network (i.e., the number of convolutional and fully connected layers), see \cite{Ref043}. These specific architectures were selected to span a representative range of model capacities, allowing for a systematic investigation of how increasing backbone depth affects crack-segmentation performance
and overfitting tendencies.}

\subsubsection{The critical role of skip connections}
\label{skip_connections}
A hallmark of the U-Net architecture is its use of long-range skip connections that link the encoder and decoder paths. After each up-sampling step in the decoder, the resulting feature map is concatenated with the corresponding feature map from the encoder at the same spatial level. This is a critical mechanism for achieving precise segmentation.

During the down-sampling process in the encoder, fine-grained spatial information (the "where") is progressively lost in favor of high-level semantic information (the "what"). While this semantic information is crucial for understanding the overall content of the image, the precise localization needed to draw sharp crack boundaries requires the lost spatial details. The skip connections re-introduce these high-resolution features directly to the decoder. By combining the "what" from the up-sampled decoder path with the "where" from the encoder's skip connections, the network can make highly informed, pixel-perfect predictions, resulting in clean and accurately delineated segmentation masks. Without these connections, the decoder would rely solely on the low-resolution bottleneck features, leading to blurry and imprecise outputs.

\subsection{Dataset generation}
\label{Dataset}

\textcolor{black}{All high-fidelity simulations are carried out in \texttt{ABAQUS} using a custom user-defined element (UEL) implementing the coupled electromechanical phase-field formulation. The nodal field outputs of the phase-field variable $\mathfrak{d}$ and electric potential $\phi$ are exported and post-processed into structured grayscale images on a uniform grid. These images form a physically grounded dataset encapsulating geometrical configurations, electric potential, and fracture damage under prescribed mechanical and electrical loading scenarios. The dataset serves as the interface between the physics-based model and the data-driven stage.}

\textcolor{black}{The deep learning workflow is implemented in a standalone Python environment using \textsc{TensorFlow/Keras}. The geometric configuration of the plate--specifically the binary distribution of holes--is used as the primary input for the CNN models. The exported physical field images from the FEM simulations (phase-field and electric potential) serve as the ground-truth targets to guide the network during the training and evaluation phases. No direct coupling with \texttt{ABAQUS} is required during prediction; surrogate modeling is performed offline using the pre-generated dataset. In the following sections, the simulation framework, data structure, and augmentation strategies used to generate this diverse, geometry-to-fracture mapping dataset are detailed.}

\subsubsection{Simulation framework}
\label{Simulation}

The dataset was generated using \texttt{ABAQUS} in combination with a custom-developed UEL subroutine that implements a fully coupled electromechanical phase-field fracture formulation. This framework simultaneously solves for the displacement field, electric potential, and phase-field fracture variable, allowing accurate modeling of electromechanically induced fracture in heterogeneous media.

The material system considered in this study is a boehmite nanoparticle-reinforced epoxy composite, designed to emulate the electromechanical behavior of dielectric nanocomposites used in soft sensing and actuation. The relevant material parameters employed in the simulations are summarized in Table~\ref{tab:material-properties}.

\begin{table}[htbp]
  \centering
  \caption{Material properties of the boehmite nanoparticle-reinforced polymer composite used in simulations, see \cite{Ref044,Ref045}.}
  \label{tab:material-properties}
  
  \begin{tabular}{llc}
	\hline
    Property & Symbol & Value \\
	\hline
	\hline
    Young's modulus & $E$ & 3.75 GPa \\
    Poisson's ratio & $\nu$ & 0.33 \\
    Dielectric permittivity & $\epsilon$ & $4.9 \times 10^{-11}$ \\
    Fracture toughness & $\mathcal{G}_c$ & 165 J/m$^2$ \\
    Degradation threshold & $k$ & $1 \times 10^{-4}$ \\
    Degradation exponent & $n$ & 2.0 \\
    \hline
  \end{tabular}
\end{table}

\textcolor{black}{The simulated specimen consists of a square plate measuring $1 \text{ mm} \times 1 \text{ mm}$ with a thickness of $0.05 \text{ mm}$, representative of a thin-film nanocomposite material. To introduce geometric variability and simulate defect-induced failure mechanisms, each model incorporates 10 to 15 randomly distributed circular holes with diameters ranging from $0.02 \text{ mm}$ to $0.06 \text{ mm}$. This stochastic geometric setup serves as a geometry-agnostic benchmark designed to generate a statistically rich and diverse dataset of crack-defect interactions. By varying the spatial configuration of these holes, we ensure the deep learning framework learns the underlying physics of electromechanical field perturbations rather than memorizing a specific structural shape, thereby enhancing its generalizability.} 

A mesh convergence study was conducted to ensure the quality and accuracy of the finite element discretization prior to running the full simulation set. Hole placement and sizing were automated using a custom Python script integrated within \texttt{ABAQUS}, enabling efficient generation of spatially randomized but reproducible geometries across 10,000 simulation instances.

The computational mesh was designed to balance accuracy and efficiency via a non-uniform meshing strategy. A coarser mesh with an element size of approximately 0.02 mm was used in regions distant from stress concentrations, while a refined mesh of 0.005 mm was applied near the holes to capture local gradients in stress and electric fields. This refinement was especially important for resolving fracture localization and crack initiation around the defects. A characteristic length scale of 0.075 mm was used in accordance with the requirements of the phase-field fracture model to ensure mesh-objective crack resolution.

\textcolor{black}{In terms of loading, a displacement-controlled test was simulated by prescribing a total displacement of 0.15 mm at the right edge, while fixing the left edge. For the electromechanical coupling, Dirichlet boundary conditions were applied with a potential difference of 1 V across the specimen. These boundary conditions were simplified to isolate the fundamental electromechanical coupling and prevent the dataset from being biased toward a specific structural geometry.}

\textcolor{black}{This configuration is highly relevant to Structural Health Monitoring (SHM) applications. In practical applications, the electric potential represents a physically measurable quantity via electrodes; our model demonstrates that these measurable field perturbations act as a unique 'signature' that can be mapped directly to internal fracture states. To prevent numerical artifacts, all holes were positioned at least 0.1 mm from the edges, ensuring crack initiation is driven solely by internal defect interactions.}

\begin{figure}
\begin{center}
\includegraphics[width=0.7\linewidth]{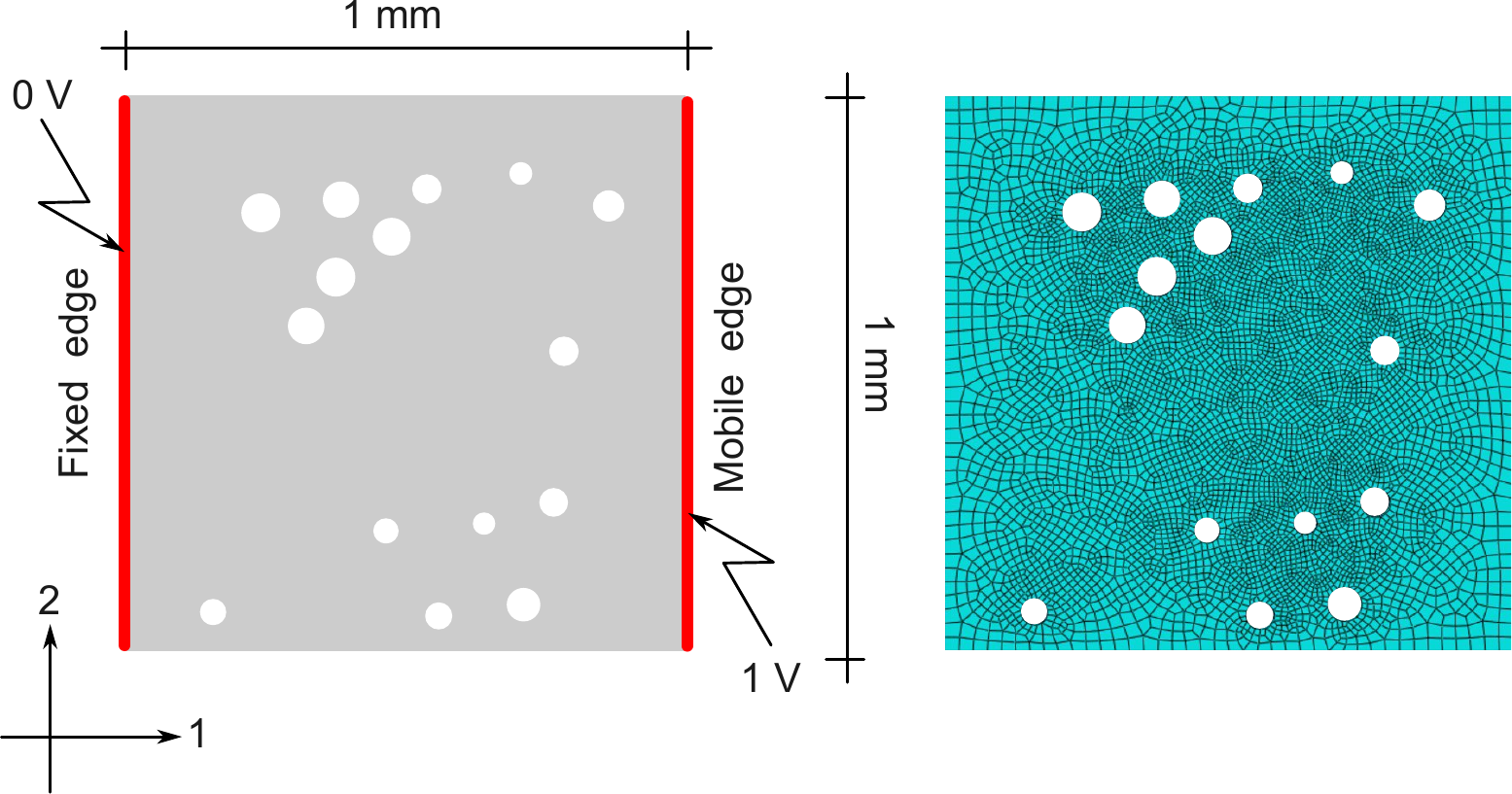}
\caption{(Left) Model geometry and boundary conditions. (Right) finite element (FE) mesh.}
\label{figure1}
\end{center}
\end{figure}

A schematic representation of the model geometry and applied boundary conditions (left) and the corresponding finite element mesh (right) are shown in Figure~\ref{figure1}.

This carefully designed simulation setup--combining geometric randomness, electromechanical coupling, and adaptive finite element discretization--was essential for generating a high-quality, diverse dataset. It provides the basis for analyzing the interplay between structural defects, mechanical fracture, and electric field perturbations in dielectric  nanocomposite materials.

\subsubsection{Dataset composition}

The final dataset comprises 10,000 distinct high-fidelity simulations, each representing a unique electromechanical fracture scenario. A key source of variability in the simulations stems from the random spatial distribution of circular holes within the plate. This geometric randomness played a crucial role in influencing crack initiation, propagation paths, and stress localization. As a result, every simulation produced a distinct crack path, enhancing the diversity and generalizability of the dataset. From a computational standpoint, each simulation was executed using 32 CPUs and required at most 15 seconds.

The inclusion of these randomly placed defects led to a wide range of stress states and electrical responses across the simulation domain, providing a comprehensive understanding of the material’s behavior under varying mechanical and electrical loading conditions. This variability is essential for developing robust data-driven models capable of capturing the complex interaction between material heterogeneity and fracture evolution.

Upon completion of each simulation, the results were post-processed to extract two primary image-based fields: (i) the phase-field fracture variable ($\mathfrak{d}$), which captures the spatial evolution of fracture and (ii) the electric potential distribution ($\phi$) across the plate. These fields were exported as high-resolution grayscale, enabling direct application of deep learning algorithms commonly used in semantic segmentation and image-to-image translation. 

\begin{figure}[]
\begin{center}
\includegraphics[width=0.8\linewidth]{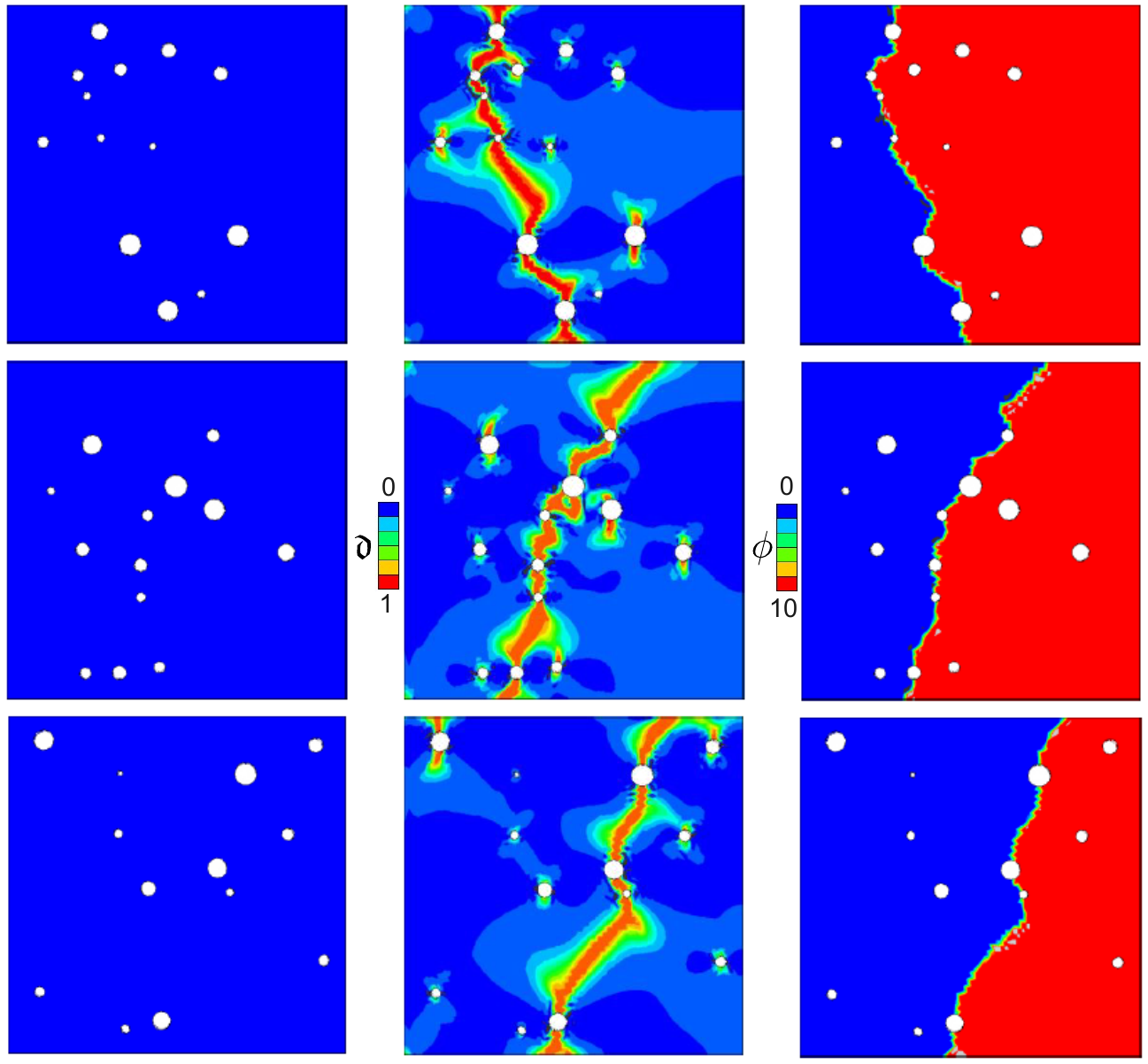}
\caption{Representative simulations -- (Left) Reference/original configuration, (Middle) Phase-field crack path, and (Right) Electric potential crack path. \textcolor{black}{Note: These plots are shown as standard false-color (RGB) contours for illustrative clarity; the actual data utilized for the deep learning framework are single-channel grayscale images derived from these distributions.}}
\label{figure2}
\end{center}
\end{figure}

The phase-field images provided detailed spatial maps of crack initiation, growth, coalescence, and branching. These images serve as direct visualizations of the fracture process, see Figure~\ref{figure2} (middle). By analyzing the progression of $\mathfrak{d}$, insights into fracture localization and crack morphology were obtained, which are crucial for understanding fracture mechanics in smart materials.

In parallel, the electric potential images offered complementary insights into how fracture affects the material’s electromechanical behavior. As shown in Figure~\ref{figure2} (right), the presence of cracks introduces measurable distortions in the potential field. These distortions often correspond to high-stress regions and crack interfaces, enhancing contrast and facilitating crack detection through electric field variation. This coupling between electrical and mechanical fields provides a unique vantage point for observing the influence of mechanical fracture on functional electrical performance.

Together, the phase-field and electric potential images form a rich visual dataset that is not only useful for physical interpretation but also serves as foundational training and validation data for machine learning models.

\textcolor{black}{To provide a clear overview of the framework’s implementation, the process is structured into distinct offline and online stages. This separation ensures that the computational complexity of the fully coupled electromechanical simulations is confined to the data-generation phase, while the predictive phase remains computationally lightweight. Algorithm~\ref{alg:workflow} outlines the step-by-step workflow, from the initial physics-based modeling to the final surrogate-based inference. Crucially, while the training process is informed by coupled physical signatures--specifically the phase-field and electric potential data--the online deployment requires only the geometric configuration as input, enabling near-instantaneous fracture assessment without the need for live field calculations.}

\begin{algorithm}[H]
	\caption{\textcolor{black}{Implementation of the hybrid electromechanical DL framework.}}
	\label{alg:workflow}
	\begin{algorithmic}[1]
		\STATE \textbf{Phase I: Offline high-fidelity data generation}
		\STATE \textit{Geometric initialization:} Define $N$ samples of the domain with varying defect (hole) configurations.
		\STATE \textit{Coupled PF-simulation:} Solve the governing electromechanical equations for each sample to obtain the final fracture state.
		\STATE \textit{Data extraction:} Export the initial geometric configuration ($X$), along with the resulting phase-field ($\phi$) and electric potential field, highlighting the final crack path.
		\STATE \textbf{Phase II: Training and physics-based learning}
		\STATE \textit{Data normalization:} Scale physical fields $\phi$ and $\mathfrak{d}$ to the range $[0, 1]$ for numerical stability.
		\STATE \textit{CNN training:} Train the ResNet-U-Net to map geometric features to the final fracture path.
		\STATE \textit{Feature supervision:} Compare training strategies supervised by phase-field targets vs. those informed by global electric potential gradients.
		
		\STATE \textbf{Phase III: Predictive inference (Deployment)}
		\STATE \textit{Input reception:} Input a new geometric configuration image into the trained network.
		\STATE \textit{Surrogate execution:} Perform a forward pass through the CNN (bypassing the iterative FEM solver).
		\STATE \textit{Result output:} Generate the predicted crack path mask for instantaneous assessment.
	\end{algorithmic}
\end{algorithm}

\section{Results and discussion}
\label{Results}

This section presents an in-depth evaluation of the proposed hybrid framework integrating phase-field fracture mechanics with convolutional neural networks. We analyze the models’ effectiveness in learning and generalizing fracture characteristics in dielectric nanocomposites under electromechanical loading. The results are organized into five subsections, covering the evaluation metrics, training-validation trends, model performance with phase-field and electric potential data, and a comparative analysis between the two.

\subsection{Evaluation metrics}
\label{metrics}

To quantitatively assess the predictive accuracy of the CNN models across different modalities--namely, phase-field fracture and electric potential images--a set of core evaluation metrics was employed. The F1-score was used to balance precision and recall, making it particularly effective for binary segmentation tasks where there is a class imbalance between fractured and unfractured regions. Intersection over Union (IoU) was adopted to quantify the spatial overlap between the predicted crack regions and the ground truth, providing a robust measure of localization performance. To evaluate model convergence, a combination of binary cross-entropy and dice loss functions was used, with lower values indicating closer alignment between the predicted and actual segmentation maps. Additionally, validation metrics--including validation F1-score (Val F1), validation IoU (Val IoU), and validation loss (Val Loss)--were tracked to assess the models' generalization capabilities on unseen data. Finally, training time was recorded to evaluate computational efficiency, which is critical for practical deployment in real-time or resource-constrained environments. All metrics were computed per epoch and averaged across the training and validation datasets to ensure statistical consistency and robust model assessment.

The F1-score is defined as:
\begin{equation}
    \text{F1} = 2 \cdot \frac{\text{Precision} \cdot \text{Recall}}{\text{Precision} + \text{Recall}},
    \label{Eq31}
\end{equation}

and the IoU is given by:
\begin{equation}
    \text{IoU} = \frac{\text{True positives}}{\text{True positives} + \text{False positives} + \text{False negatives}}.
    \label{Eq32}
\end{equation}

\subsection{Training and validation}
\label{Training}

The training and validation phases were conducted using CNNs based on the ResNet-U-Net architecture. Specifically, four distinct backbone configurations were evaluated--ResNet34, ResNet50, ResNet101, and ResNet152--each integrated into a symmetric U-Net-style decoder framework. These encoder-decoder models were designed to facilitate high-resolution semantic segmentation of crack propagation paths in electromechanically loaded nanocomposite plates.

\textcolor{black}{The networks were trained separately on two data modalities derived from physics-based FE simulations: (i) the phase-field scalar field ($\mathfrak{d}$) and (ii) the electric potential ($\phi$). Each dataset comprised 10,000 unique single-channel images of size $572 \times 572$ pixels, representing 2D slices of simulated specimens with randomly distributed defects. Since the finite element simulations utilize an unstructured mesh to capture the electromechanical response, a post-processing rasterization step was performed to interpolate the nodal field data onto this structured Cartesian grid to ensure compatibility with the CNN architecture.} 

\textcolor{black}{A critical distinction between these modalities is their numerical range; while the phase-field variable $\mathfrak{d}$ is inherently bounded in $[0, 1]$, the raw electric potential $\phi$ is physically dependent on the applied boundary conditions and material properties. To ensure architectural consistency and training stability, both modalities were normalized to a common range of $[0, 1]$ before being passed to the ResNet-U-Net. Specifically, the electric potential fields were linearly rescaled for each sample such that the local minimum and maximum values map to 0 and 1, respectively. This min-max normalization makes the network invariant to absolute voltage levels and ensures that the \textcolor{black}{training} statistics are consistent across all training samples. While finite element results are often visualized as RGB contour plots for illustrative purposes (see Figure~\ref{figure2}), the actual \textcolor{black}{data} used for training and inference are strictly normalized grayscale images to optimize computational efficiency. The crack paths captured via phase-field or electric potential gradients served as the ground-truth labels for supervised segmentation learning.}

All networks were implemented using TensorFlow and trained on an NVIDIA RTX 3090 GPU with 24 GB of memory. The training process employed the Adam optimizer with an initial learning rate of \(1 \times 10^{-4}\), using \(\beta_1 = 0.9\) and \(\beta_2 = 0.999\) to control the decay rates for the moment estimates. An adaptive learning rate scheduler was used to reduce the learning rate by a factor of 0.1 if the validation loss plateaued. To prevent overfitting, early stopping was applied. The loss function was a composite of 50\% Dice loss and 50\% binary cross-entropy loss, allowing the model to balance region-level accuracy with pixel-wise classification. Training was conducted over a maximum of 100 epochs, subject to early stopping criteria, using a batch size of 24. To enhance generalization, data augmentation was applied in the form of random rotations (\(\pm 15^\circ\)), horizontal and vertical flips, and additive Gaussian noise with standard deviation \(\sigma = 0.05\), see \cite{Ref047,Ref048}. The dataset was randomly partitioned into 80\% training, 10\% validation, and  10\% testing subsets using stratified sampling to maintain label distribution.  

\begin{figure}[]
\begin{center}
\includegraphics[width=0.99\linewidth]{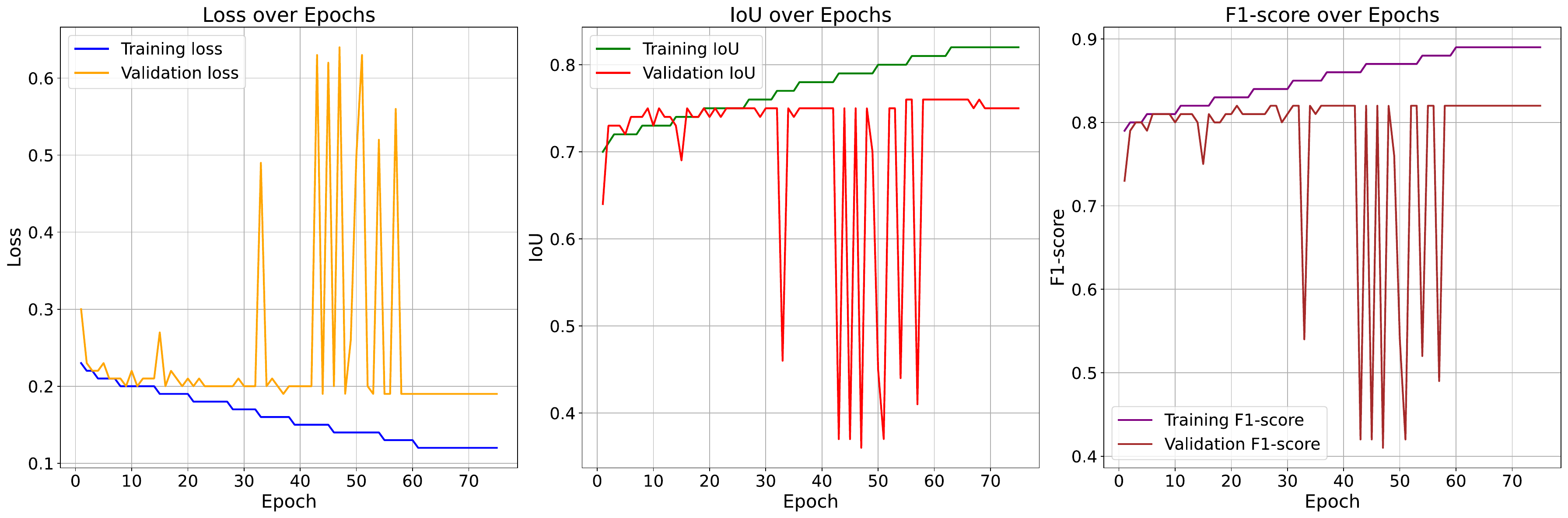}
\caption{Training and validation metrics for the ResNet34-based U-Net model on phase-field data -- (Left) Loss, (Middle) IoU, and (Right) F1-score.}
\label{figure3}
\end{center}
\end{figure}

As a representative example, the training and validation metrics for the ResNet34-based U-Net model are shown in Figure~\ref{figure3} for the phase-field data. Initially, the model yielded an F1-score of 0.75 and an IoU of 0.67, with a training loss of 0.27. Over the course of 75 epochs, the model achieved an F1-score of 0.89, IoU of 0.82, and a reduced training loss of 0.12, with corresponding validation metrics stabilizing at 0.82, 0.75, and 0.19 respectively.

\begin{figure}[]
\begin{center}
\includegraphics[width=0.99\linewidth]{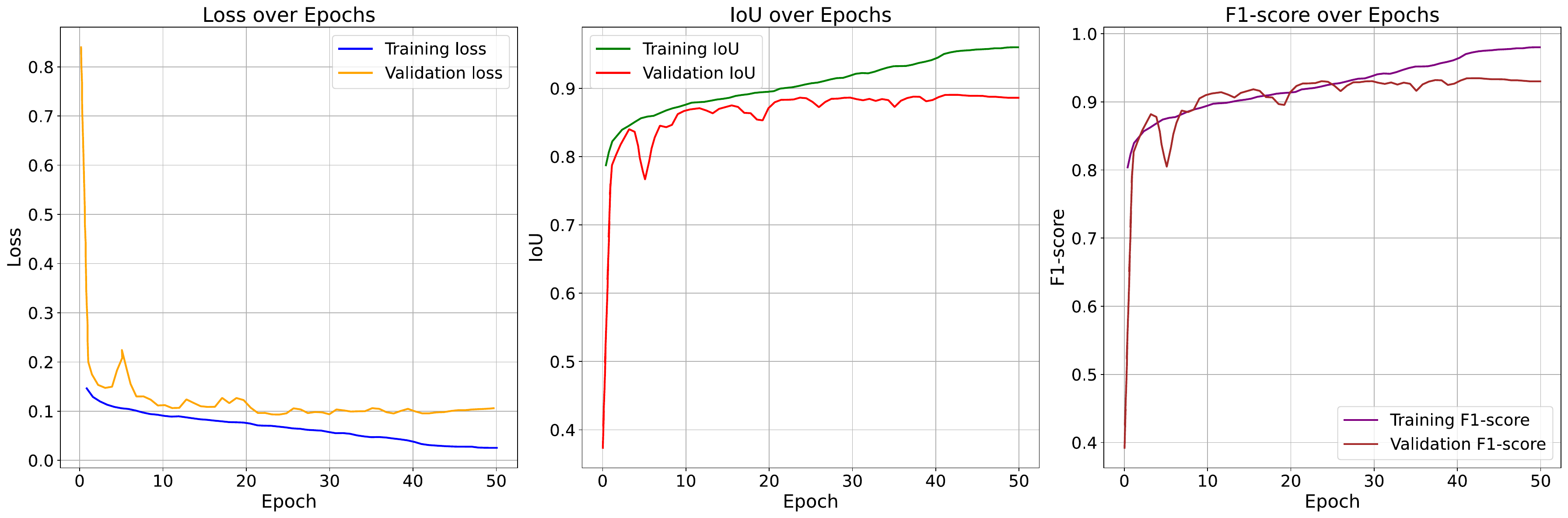}
\caption{Training and validation metrics for the ResNet34-based U-Net model on electric potential data -- (Left) Loss, (Middle) IoU, and (Right) F1-score.}
\label{figure4}
\end{center}
\end{figure}

As for the electric potential-based training, convergence occurred earlier--within approximately 50 epochs, see Figure~\ref{figure4}. The ResNet34 model reached a training F1-score of 0.98, IoU of 0.96, and a training loss as low as 0.027. Validation scores also remained high, at F1-score = 0.93 and IoU = 0.89, with a loss of 0.109, indicating strong generalization capabilities.

The consistent reduction in loss and concurrent increase in F1 and IoU scores for both data types confirms that the networks were effectively learning spatial correlations and developing strong crack segmentation capabilities. Importantly, the model trained using electric potential \textcolor{black}{signatures} exhibited faster convergence and superior validation metrics, likely due to the smoother and more continuous field structure in electric potential distributions, which may offer richer gradient information to CNNs compared to localized phase-field \textcolor{black}{data}.

The training process incorporated several regularization strategies to enhance model robustness and generalization. Dropout layers with a rate of 0.2 were introduced in the bottleneck region of the decoder path to reduce co-adaptation of neurons. Additionally, batch normalization was applied after every convolutional layer to stabilize gradient flow and accelerate convergence. The diversity introduced through data augmentation--including geometric transformations and noise injection--further improved the network's resilience to variability in crack morphology and positioning. Throughout training, no signs of overfitting were observed. The validation loss closely followed the training loss across all network architectures, with ResNet34 and ResNet152 demonstrating an optimal balance between model complexity and stable generalization performance.

\subsection{Performance with phase-feld data}
\label{phase-feld data}
Phase-field data represent a scalar field $\mathfrak{d}$ that models the progressive degradation of material stiffness, ranging from 0 (intact) to 1 (fully fractured). In the electromechanical phase-field model, $\mathfrak{d}$ evolves based on energy minimization principles and provides spatially localized crack information. Given its high fidelity and sharp contrast near fracture zones, it has traditionally been used as a benchmark for crack path identification in fracture mechanics.

Herein, the CNN models were trained to learn and segment crack paths directly from phase-field \textcolor{black}{data} using four ResNet backbones: ResNet34, ResNet50, ResNet101, and ResNet152. Each architecture was evaluated over a wide range of simulation cases with varying defect distributions and crack topologies.

Table~\ref{tab:phi_performance} summarizes the performance of the models on the phase-field dataset. \textcolor{black}{The performance metrics in this table are derived from a pixel-wise confusion matrix analysis, comparing every pixel of the CNN prediction against the ground-truth data.} Among the architectures, ResNet34 provided the best trade-off between accuracy and computational efficiency. This architecture reached a training F1-score of 0.89 and an IoU of 0.82 after 75 epochs, with a total training time of approximately 2 hours. On the validation set, ResNet34 reached an F1-score of 0.82 and an IoU of 0.75, with a final validation loss of 0.19.

In comparison, ResNet50, ResNet101, and ResNet152 required longer training times--3, 4, and 5 hours, respectively--for 50, 65, and 53 epochs. Among these three models, ResNet50 exhibited the weakest performance despite its relatively short training duration. It achieved a lower training F1-score of 0.83 and IoU of 0.74, along with the highest validation loss of 0.22 among all models. ResNet101 and ResNet152 showed moderate improvements over ResNet50 in both training and validation metrics. However, neither architecture outperformed ResNet34 in terms of the overall trade-off between segmentation accuracy and computational efficiency.

\begin{table}[htbp]
\centering
\caption{CNN Performance on phase-field data.}
\label{tab:phi_performance}
\begin{tabular}{lccccccccc}
\hline
Model & Epochs & F1-score & IoU & Loss & Val F1 & Val IoU & Val Loss & Time \\
\hline
\hline
ResNet34 & 75 & 0.89 & 0.82 & 0.12 & 0.82 & 0.75 & 0.19 & 2 hr \\
ResNet50 & 50 & 0.83 & 0.74 & 0.19 & 0.80 & 0.73 & 0.22 & 3 hr \\
ResNet101 & 65 & 0.85 & 0.77 & 0.16 & 0.81 & 0.74 & 0.20 & 4 hr \\
ResNet152 & 53 & 0.88 & 0.80 & 0.13 & 0.81 & 0.74 & 0.20 & 5 hr \\
\hline
\end{tabular}
\end{table}

The improvement in IoU over epochs is visualized in Figure~\ref{figure3} (middle), while the training/validation loss curve in Figure~\ref{figure3} (left) confirms a stable convergence pattern. Notably, the phase-field  models required a larger number of \textcolor{black}{epochs} to converge compared to the electric potential models, indicating the presence of sharper gradients and higher sensitivity in crack tip regions.

\begin{figure}[]
\begin{center}
\includegraphics[width=0.99\linewidth]{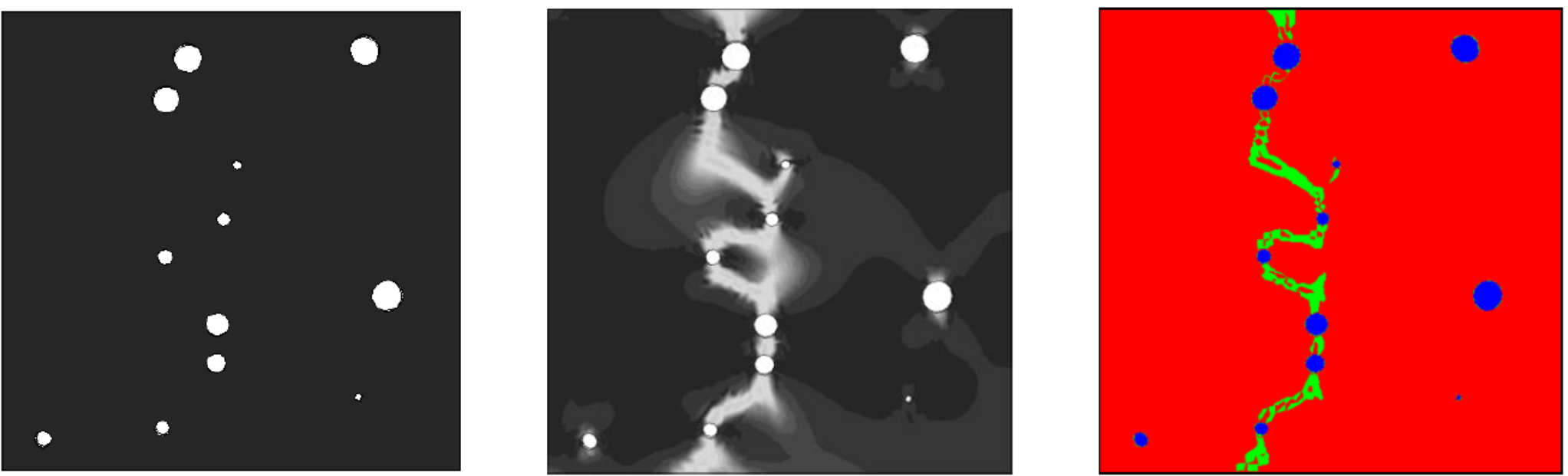}
\caption{\textcolor{black}{Case 1 -- Image preprocessing pipeline. (Left) Original input image showing the random distribution of holes, (Middle) Ground-truth crack path obtained from the phase-field damage variable in high-fidelity FE simulations, and (Right) Corresponding binary crack masks used as labels for training the deep learning model.}}
\label{figure5}
\end{center}
\end{figure}

\begin{figure}[]
\begin{center}
\includegraphics[width=0.99\linewidth]{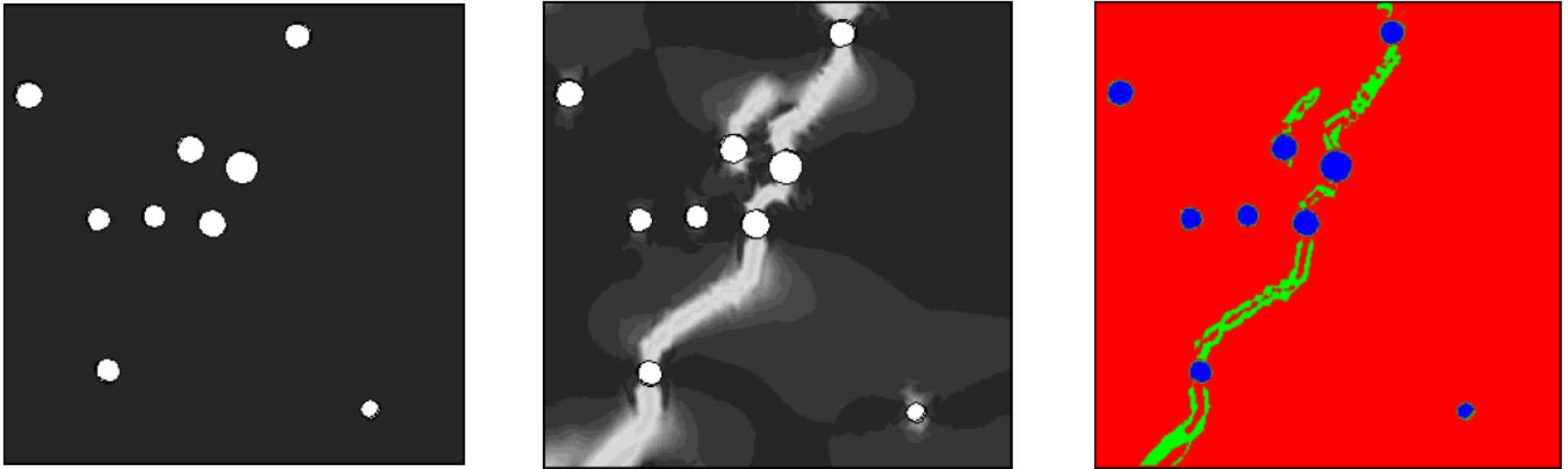}
\caption{\textcolor{black}{Case 2 -- Image preprocessing pipeline. (Left) Original input image showing the random distribution of holes, (Middle) Ground-truth crack path obtained from the phase-field damage variable in high-fidelity FE simulations, and (Right) Corresponding binary crack masks used as labels for training the deep learning model.}}
\label{figure6}
\end{center}
\end{figure}

\textcolor{black}{Figures~\ref{figure5} and \ref{figure6} illustrate two cases--Case 1 and Case 2, respectively--demonstrating the the relationship between the physical simulation and the training data. The left panels show the original input images provided to the CNN, which capture the specific geometric distribution of randomly placed circular holes. The middle panels display the actual crack paths (ground-truth) derived directly from the phase-field damage variable ($\mathfrak{d}$), while the right-most panels show the binary crack masks used as labels for CNN training.}

\begin{figure}[]
\begin{center}
\includegraphics[width=0.8\linewidth]{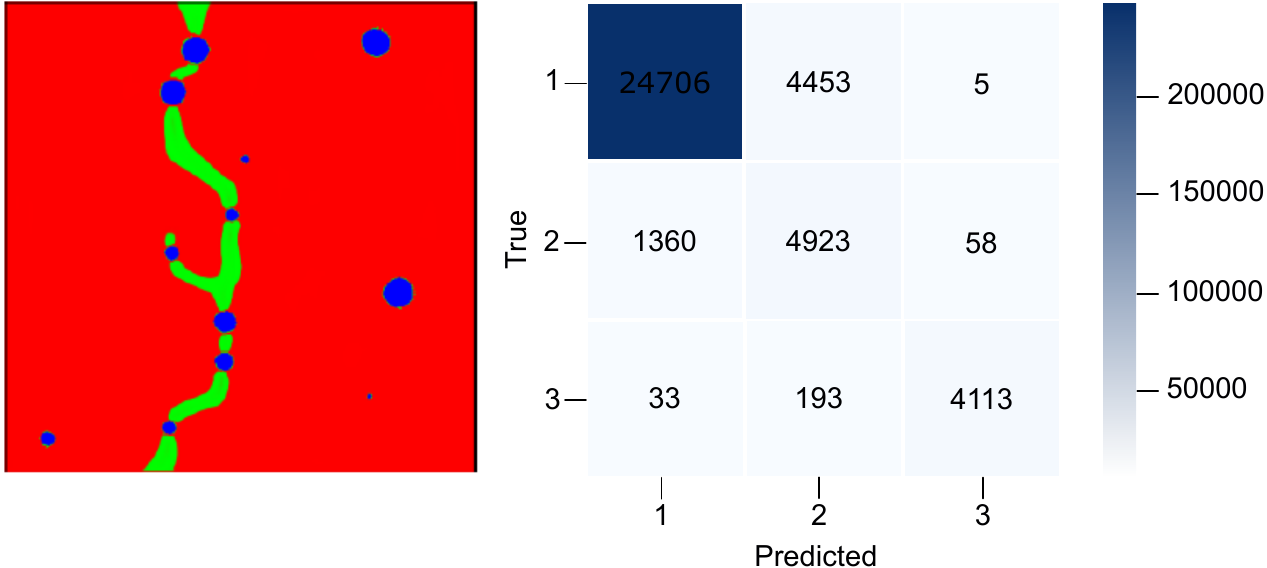}
\caption{Case 1 -- (Left) Predicted crack image for Resnet34 based on phase-field data and (Right) Confusion matrix.}
\label{figure7}
\end{center}
\end{figure}

\begin{figure}[]
\begin{center}
\includegraphics[width=0.8\linewidth]{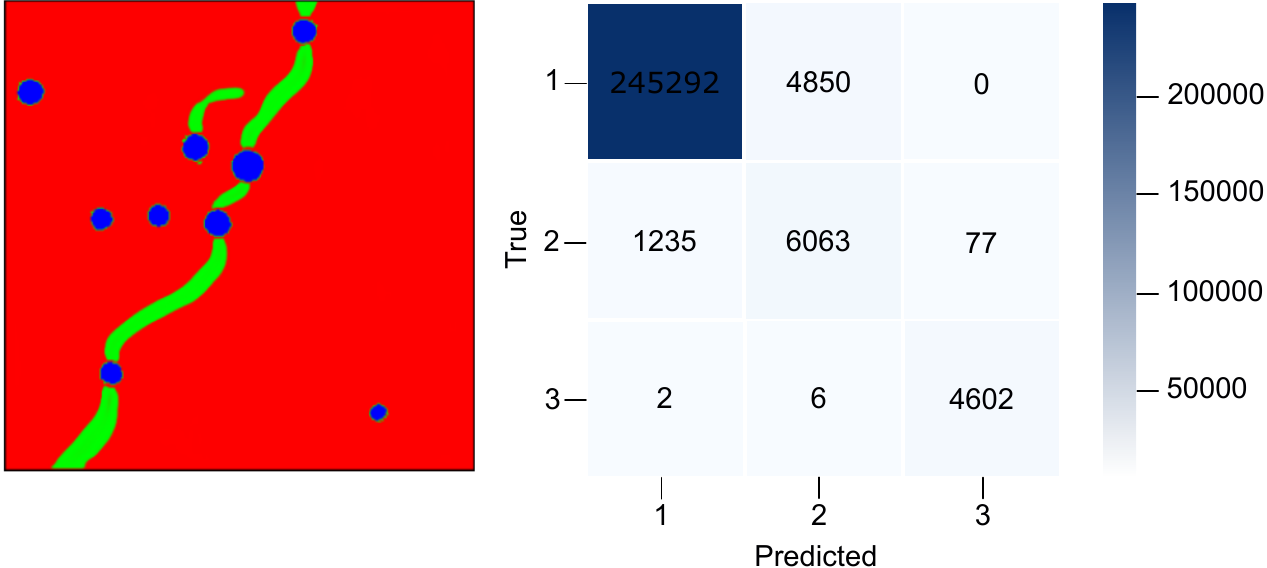}
\caption{Case 2 -- (Left) Predicted crack image for Resnet34 based on phase-field data and (Right) Confusion matrix.}
\label{figure8}
\end{center}
\end{figure}

\begin{figure}[]
\begin{center}
\includegraphics[width=0.8\linewidth]{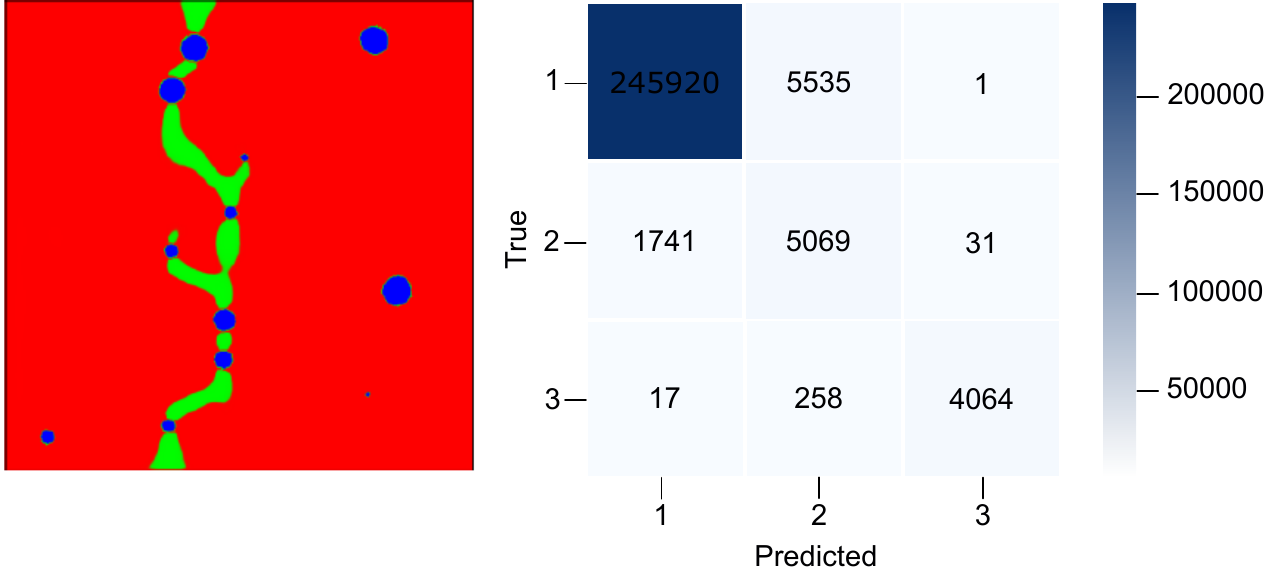}
\caption{Case 1 -- (Left) Predicted crack image for ResNet50 based on phase-field data and (Right) Confusion matrix.}
\label{figure9}
\end{center}
\end{figure}

\begin{figure}[]
\begin{center}
\includegraphics[width=0.8\linewidth]{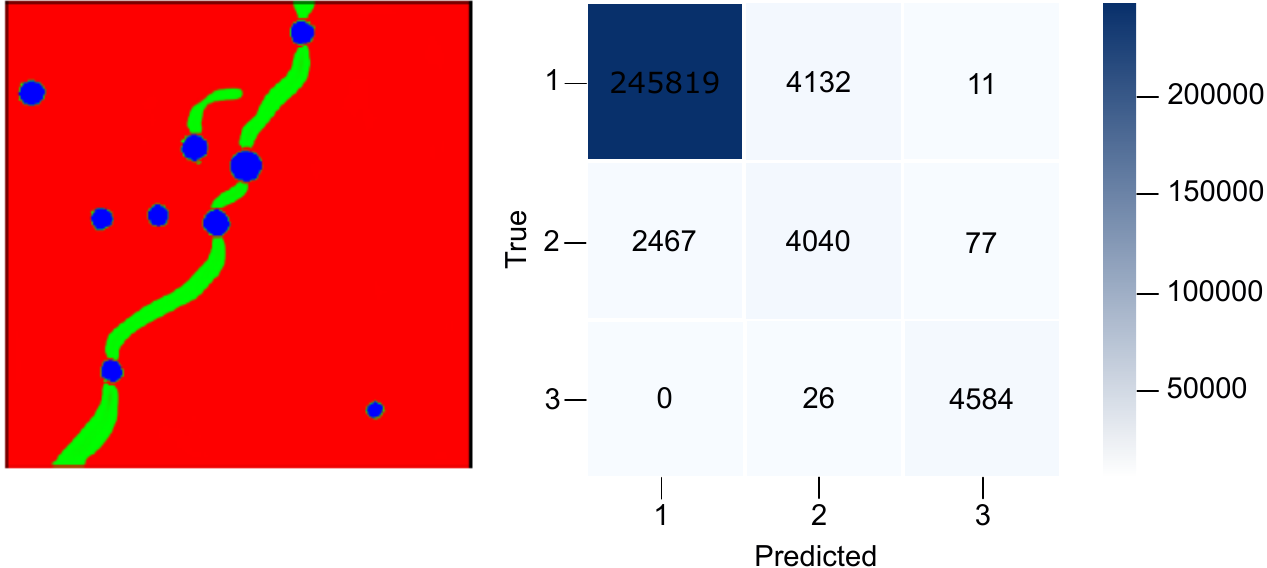}
\caption{Case 2 -- (Left) Predicted crack image for ResNet50 based on phase-field data and (Right) Confusion matrix.}
\label{figure10}
\end{center}
\end{figure}

\begin{figure}[]
\begin{center}
\includegraphics[width=0.8\linewidth]{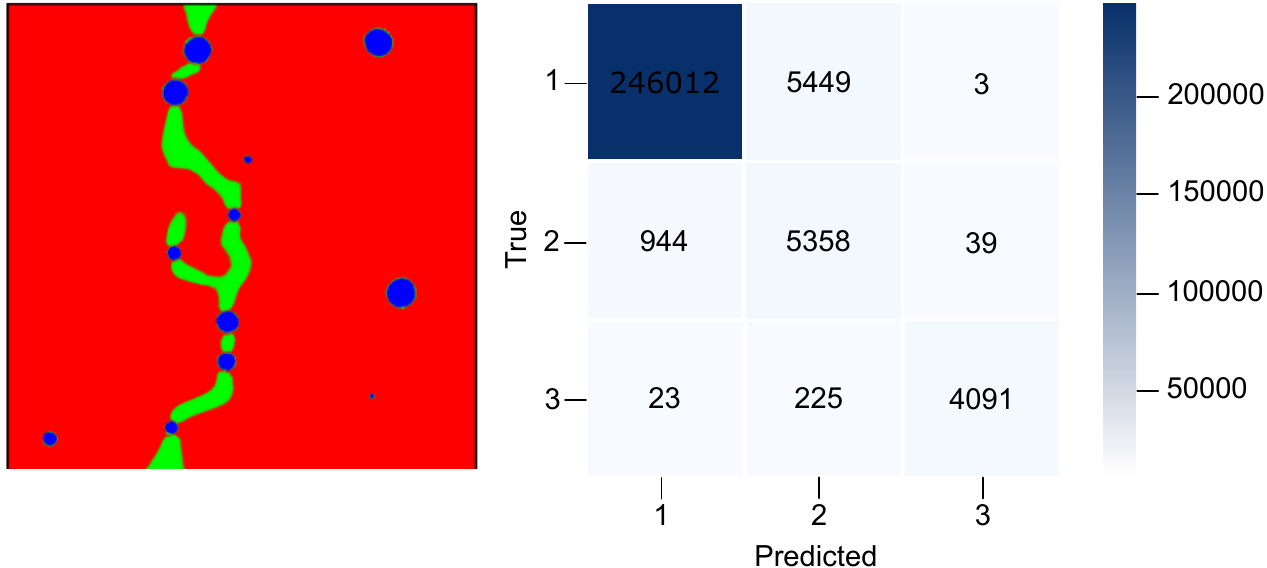}
\caption{Case 1 -- (Left) Predicted crack image for ResNet101 based on phase-field data and (Right) Confusion matrix.}
\label{figure11}
\end{center}
\end{figure}

\begin{figure}[]
\begin{center}
\includegraphics[width=0.8\linewidth]{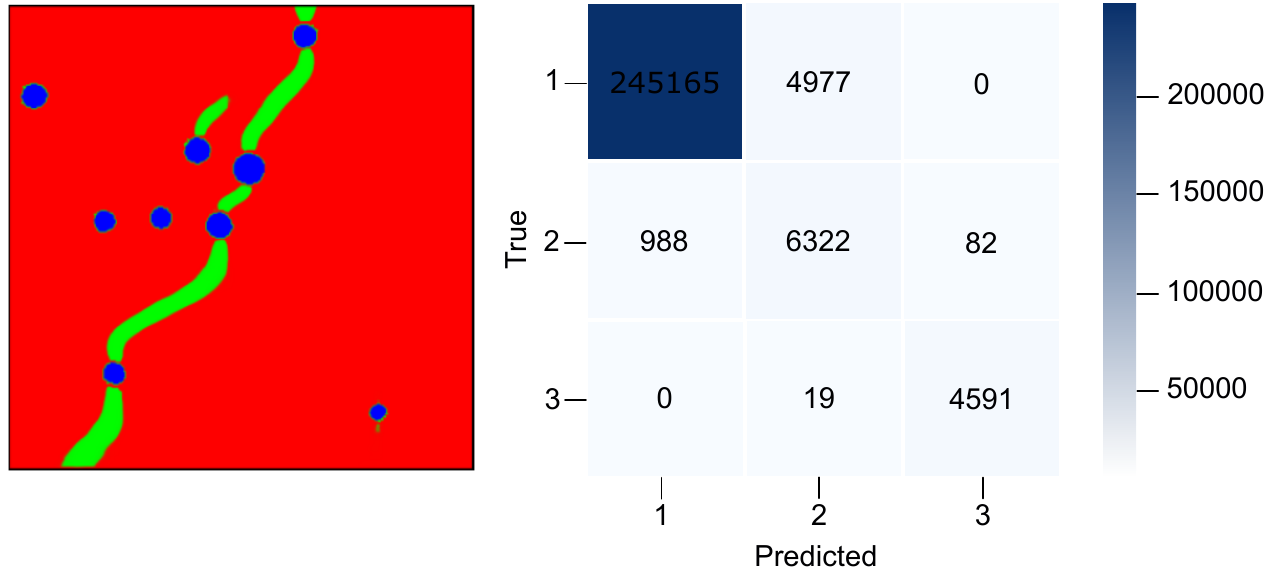}
\caption{Case 2 -- (Left) Predicted crack image for ResNet101 based on phase-field data and (Right) Confusion matrix.}
\label{figure12}
\end{center}
\end{figure}

\begin{figure}[]
\begin{center}
\includegraphics[width=0.8\linewidth]{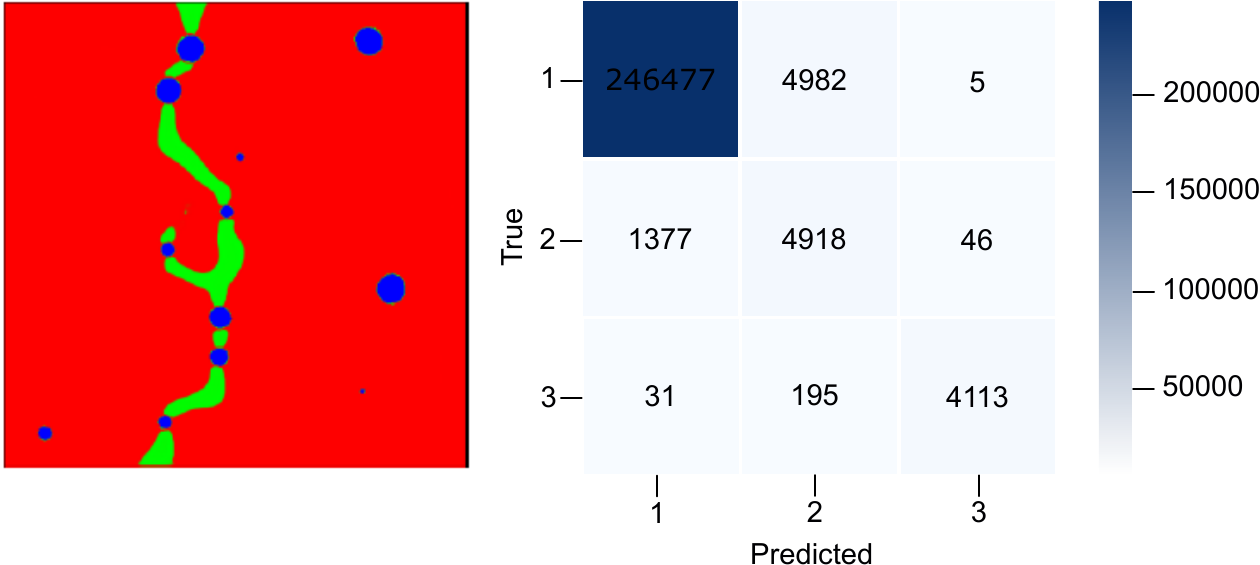}
\caption{Case 1 -- (Left) Predicted crack image for ResNet152 based on phase-field data and (Right) Confusion matrix.}
\label{figure13}
\end{center}
\end{figure}

\begin{figure}[]
\begin{center}
\includegraphics[width=0.8\linewidth]{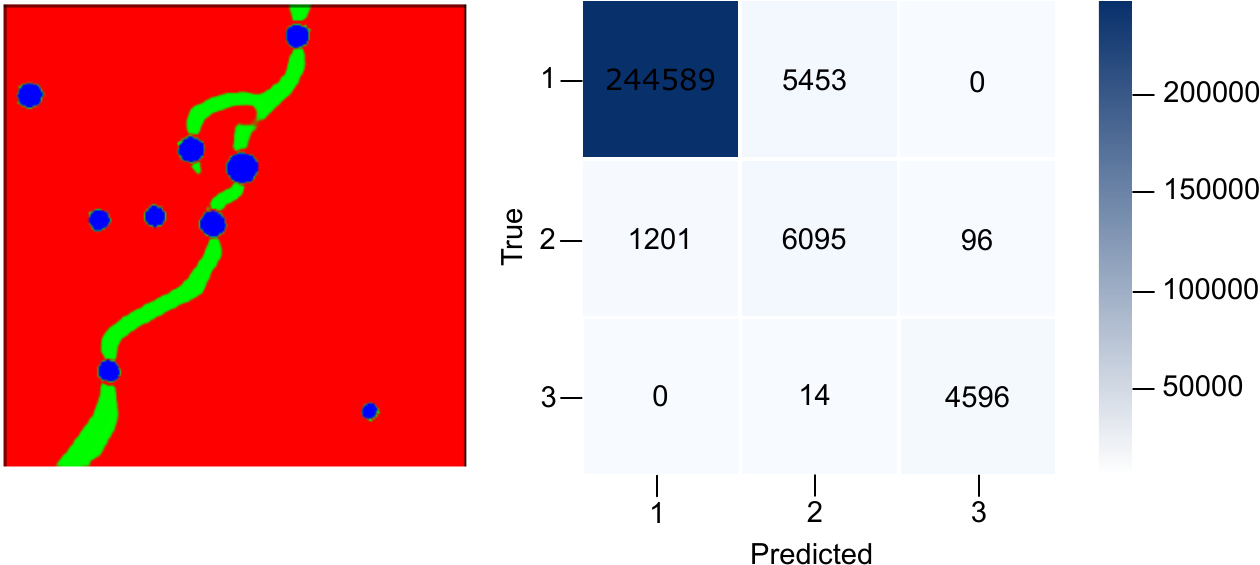}
\caption{Case 2 -- (Left) Predicted crack image for ResNet152 based on phase-field data and (Right) Confusion matrix.}
\label{figure14}
\end{center}
\end{figure}

Figures~\ref{figure7} through \ref{figure14} showcase the crack detection performance of various ResNet models for Case 1 and Case 2. Each figure consists of two components: (left) the Predicted Crack Image, illustrating the model’s ability to localize cracks, and (right) the Confusion Matrix, which quantifies the model's classification accuracy. Specifically, Figures~\ref{figure7} and \ref{figure8} correspond to ResNet34, Figures~\ref{figure9} and \ref{figure10} to ResNet50, Figures Figures~\ref{figure11} and \ref{figure12} to ResNet101, and Figures Figures~\ref{figure13} and \ref{figure14} to ResNet152. Collectively, these figures highlight differences in crack detection performance and predictive accuracy across the various ResNet architectures, offering insight into their relative effectiveness for phase-field image segmentation.

As an example, in Case 1, the predicted crack pattern closely aligns with the ground truth label, particularly in regions with branching behavior. This indicates the model's effectiveness in capturing complex crack propagation paths. However, some blurring was observed at the crack tips, which is likely attributed to limitations in the receptive field resolution of the convolutional \textcolor{black}{neural} network. The confusion matrices presented in Figures~\ref{figure7} through \ref{figure14} further support this observation, showing relatively low false positive rates across models, but highlighting the presence of false negatives in micro-cracked regions, especially where cracks are thinner and more dispersed.

Despite the physically grounded nature of phase-field imaging for fracture modeling, several inherent challenges complicate CNN-based prediction. First, the sparsity of cracks--occupying only a small fraction of the image area--introduces significant class imbalance, making it difficult for models to learn minority features effectively. Second, the discontinuous nature of crack evolution, which often involves abrupt spatial transitions, poses a challenge for convolutional filters that typically assume smooth feature variations. Third, the sensitivity of phase-field values near the decision boundary (around 0.5) can lead to classification ambiguity, especially in marginal zones between damaged and undamaged material.

Nevertheless, the CNN models demonstrated strong generalization capability and were able to predict damage propagation paths with high fidelity. 

\subsection{Performance with electric potential data}
\label{electric potential data}

Electric potential data arise from the electrostatic field solution within the electromechanical phase-field framework. In dielectric materials, crack formation disturbs the continuity of the electric field, manifesting as gradients and distortions in the potential map. Unlike the highly localized phase-field field, electric potential  images are smoother and contain distributed features, providing implicit cues about crack presence through field variation rather than direct damage encoding.

Herein, and similar to Section \ref{phase-feld data}, electric potential data were employed as a surrogate to indirectly predict crack locations. Interestingly, CNNs trained on electric potential images outperformed those trained on phase-field images, both in terms of convergence speed and segmentation accuracy.

Table~\ref{tab:v_performance} summarizes model performance on electric potential-based \textcolor{black}{strategy}. ResNet34 achieved the highest validation metrics overall, with a training F1-score of 0.98, IoU of 0.96, and a validation F1-score of 0.93 and IoU of 0.89--all while converging in just 50 epochs. ResNet152 slightly outperformed ResNet34 in final validation metrics but required 2.5$\times$ more computational time.

Figure~\ref{figure4} shows the loss and IoU trends during training, revealing a faster and more stable convergence than phase-field-based models. The electric potential-based models reached near-plateau performance in under 40 epochs, reflecting the information richness of the electric potential field for guiding crack prediction.

\begin{table}[htbp]
\centering
\caption{CNN Performance on electric potential data.}
\label{tab:v_performance}
\begin{tabular}{lccccccccc}
\hline
Model & Epochs & F1-score & IoU & Loss & Val F1 & Val IoU & Val Loss & Time \\
\hline
\hline
ResNet34 & 50 & 0.98 & 0.96 & 0.027 & 0.93 & 0.89 & 0.109 & 2 hr \\
ResNet50 & 44 & 0.96 & 0.91 & 0.064 & 0.90 & 0.84 & 0.138 & 3 hr \\
ResNet101 & 53 & 0.96 & 0.93 & 0.050 & 0.93 & 0.89 & 0.091 & 4 hr \\
ResNet152 & 50 & 0.97 & 0.94 & 0.048 & 0.94 & 0.90 & 0.084 & 5 hr \\
\hline
\end{tabular}
\end{table}

\begin{figure}[]
\begin{center}
\includegraphics[width=0.99\linewidth]{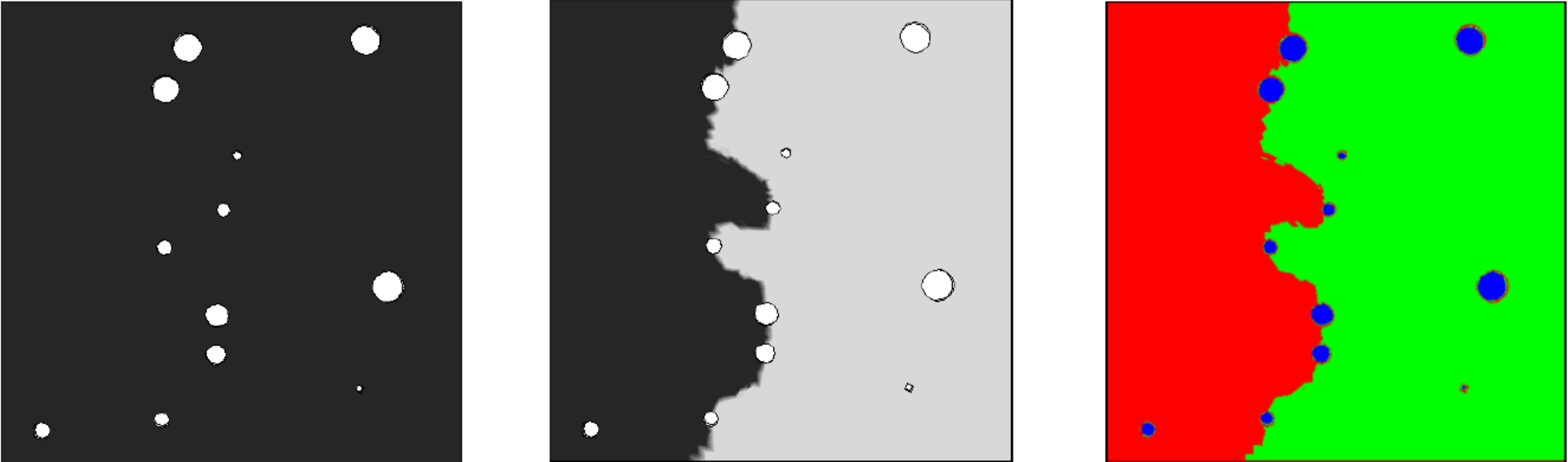}
\caption{\textcolor{black}{Case 1 -- Image preprocessing pipeline. (Left) Original input image showing the random distribution of holes, (Middle) Ground-truth crack path obtained from the electric potential field in high-fidelity FE simulations, and (Right) Corresponding binary crack masks used as labels for training the deep learning model.}}
\label{figure15}
\end{center}
\end{figure}

\begin{figure}[]
\begin{center}
\includegraphics[width=0.99\linewidth]{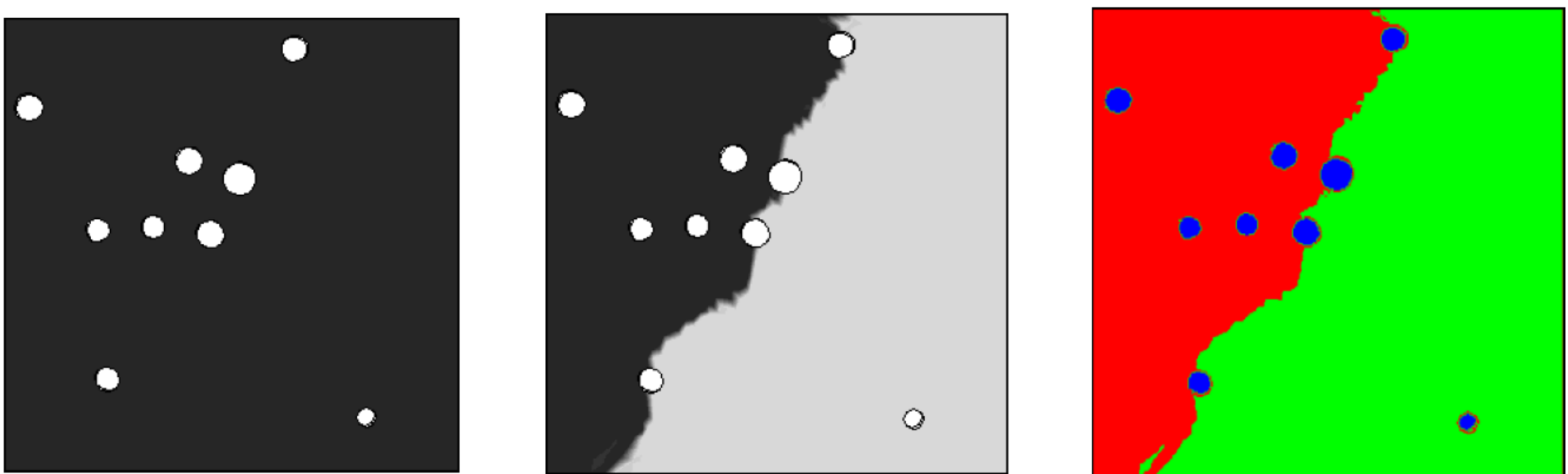}
\caption{\textcolor{black}{Case 2 -- Image preprocessing pipeline. (Left) Original input image showing the random distribution of holes, (Middle) Ground-truth crack path obtained from the electric potential field in high-fidelity FE simulations, and (Right) Corresponding binary crack masks used as labels for training the deep learning model.}}
\label{figure16}
\end{center}
\end{figure}

\begin{figure}[]
\begin{center}
\includegraphics[width=0.8\linewidth]{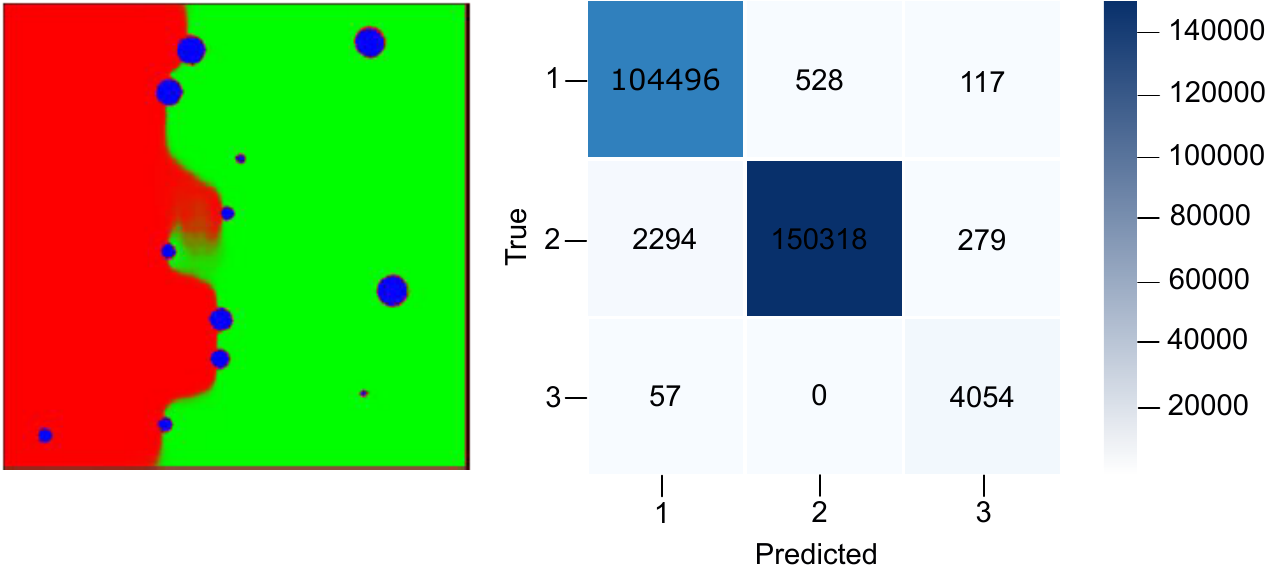}
\caption{Case 1 -- (Left) Predicted crack image for Resnet34 based on electric potential data and (Right) Confusion matrix.}
\label{figure17}
\end{center}
\end{figure}

\begin{figure}[]
\begin{center}
\includegraphics[width=0.8\linewidth]{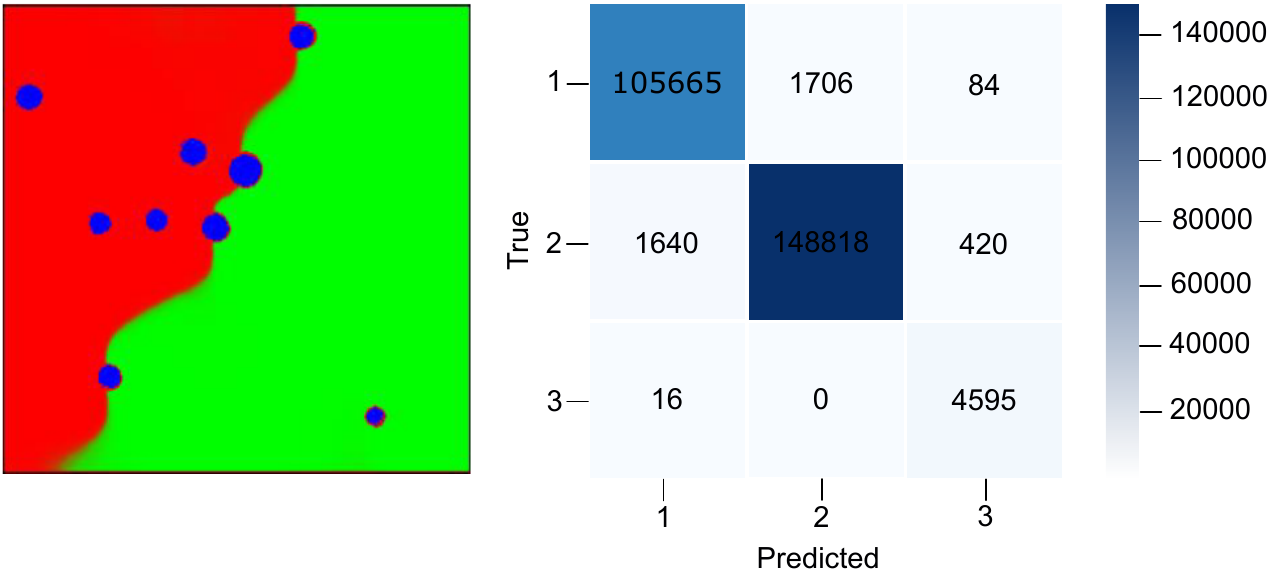}
\caption{Case 2 -- (Left) Predicted crack image for Resnet34 based on electric potential data and (Right) Confusion matrix.}
\label{figure18}
\end{center}
\end{figure}

\begin{figure}[]
\begin{center}
\includegraphics[width=0.8\linewidth]{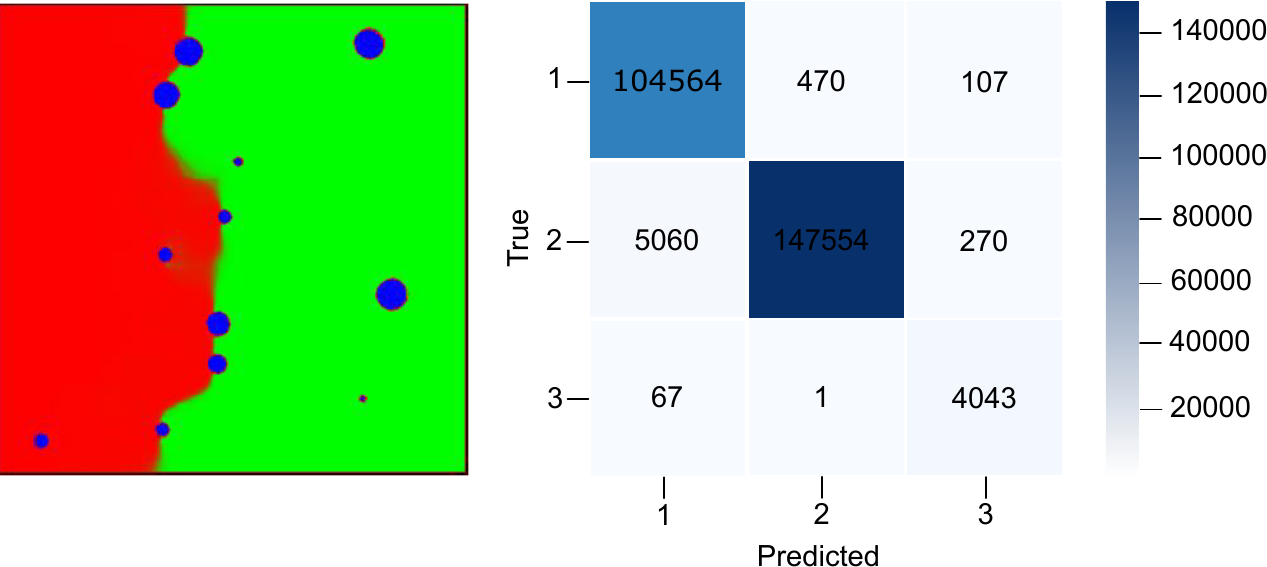}
\caption{Case 1 -- (Left) Predicted crack image for ResNet50 based on electric potential data and (Right) Confusion matrix.}
\label{figure19}
\end{center}
\end{figure}

\begin{figure}[]
\begin{center}
\includegraphics[width=0.8\linewidth]{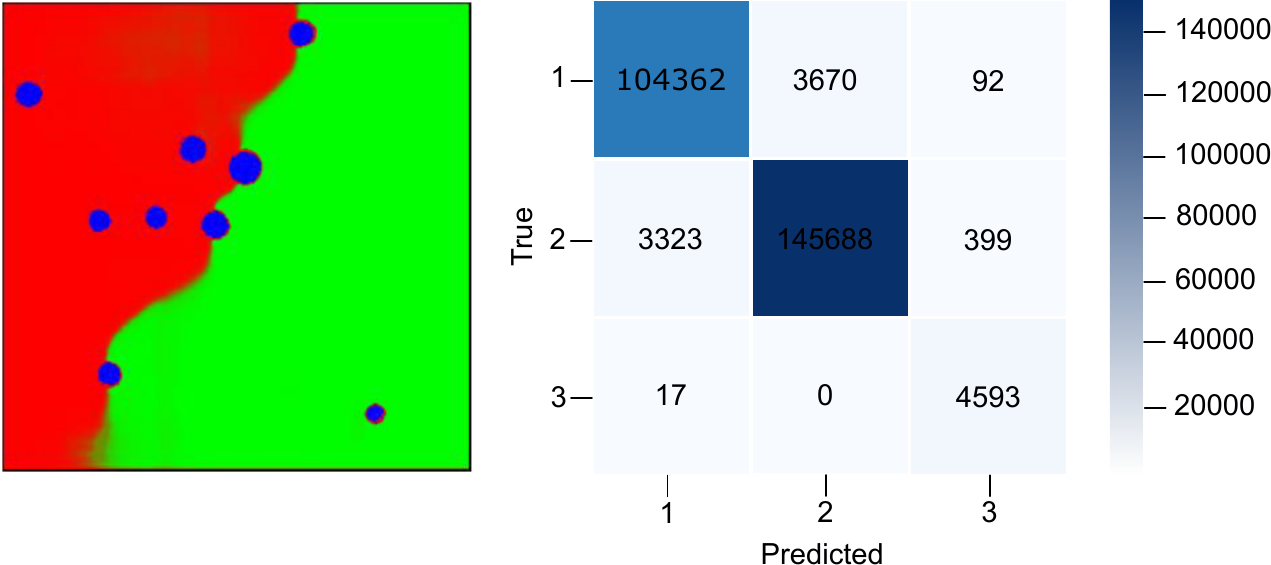}
\caption{Case 2 -- (Left) Predicted crack image for ResNet50 based on electric potential data and (Right) Confusion matrix.}
\label{figure20}
\end{center}
\end{figure}

\begin{figure}[]
\begin{center}
\includegraphics[width=0.8\linewidth]{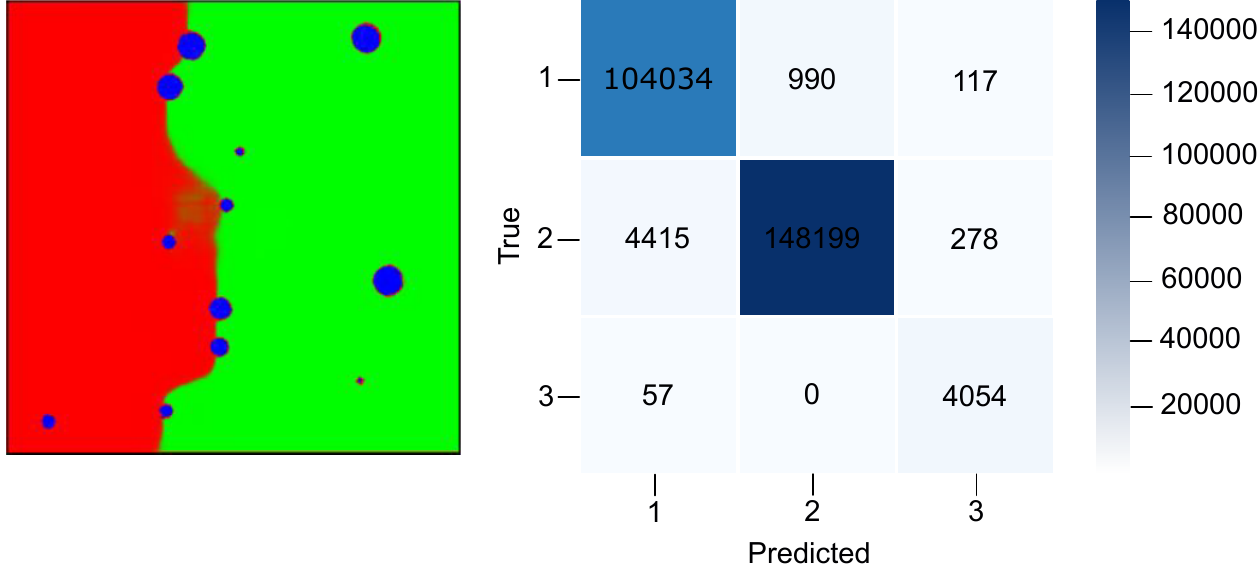}
\caption{Case 1 -- (Left) Predicted crack image for ResNet101 based on electric potentialdata and (Right) Confusion matrix.}
\label{figure21}
\end{center}
\end{figure}

\begin{figure}[]
\begin{center}
\includegraphics[width=0.8\linewidth]{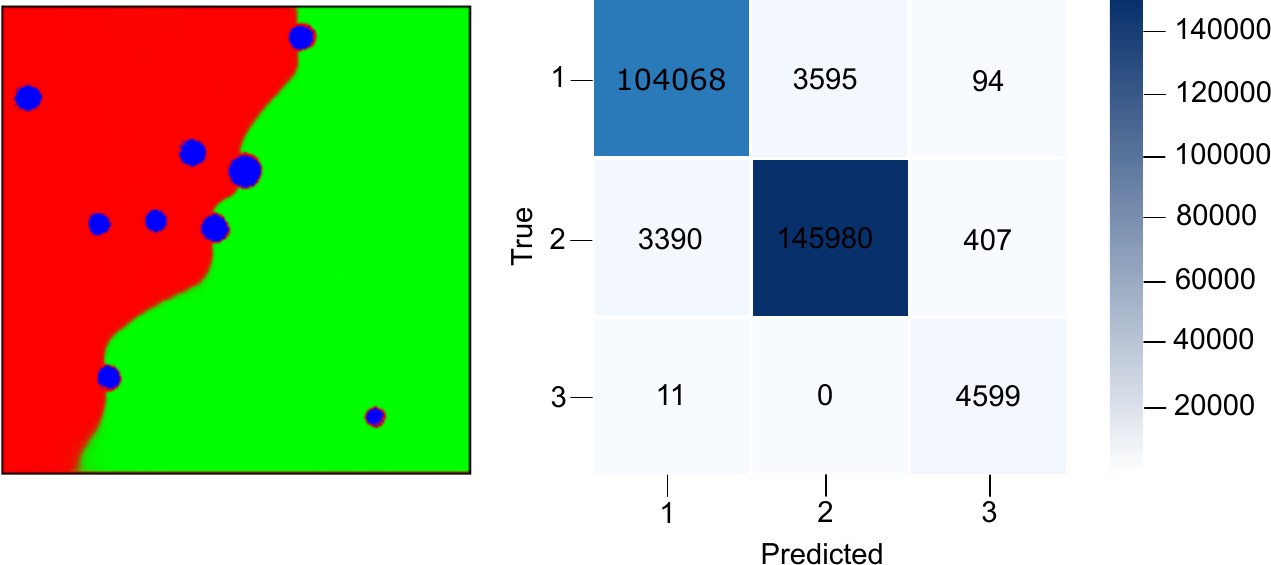}
\caption{Case 2 -- (Left) Predicted crack image for ResNet101 based on electric potential data and (Right) Confusion matrix.}
\label{figure22}
\end{center}
\end{figure}

\begin{figure}[]
\begin{center}
\includegraphics[width=0.8\linewidth]{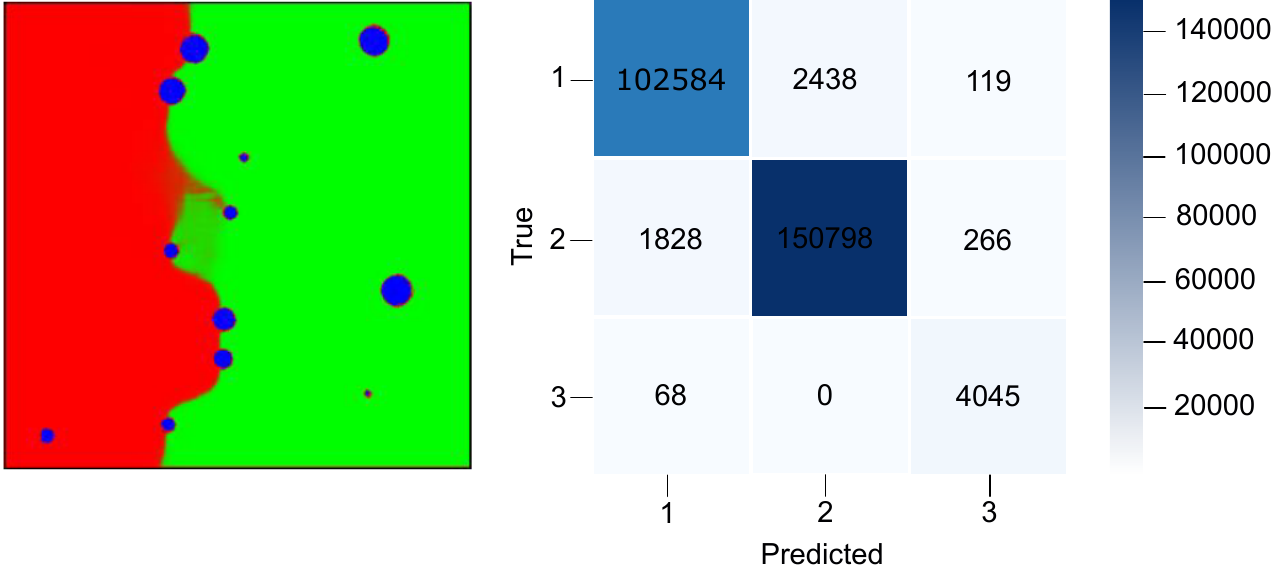}
\caption{Case 1 -- (Left) Predicted crack image for ResNet152 based on electric potential data and (Right) Confusion matrix.}
\label{figure23}
\end{center}
\end{figure}

\begin{figure}[]
\begin{center}
\includegraphics[width=0.8\linewidth]{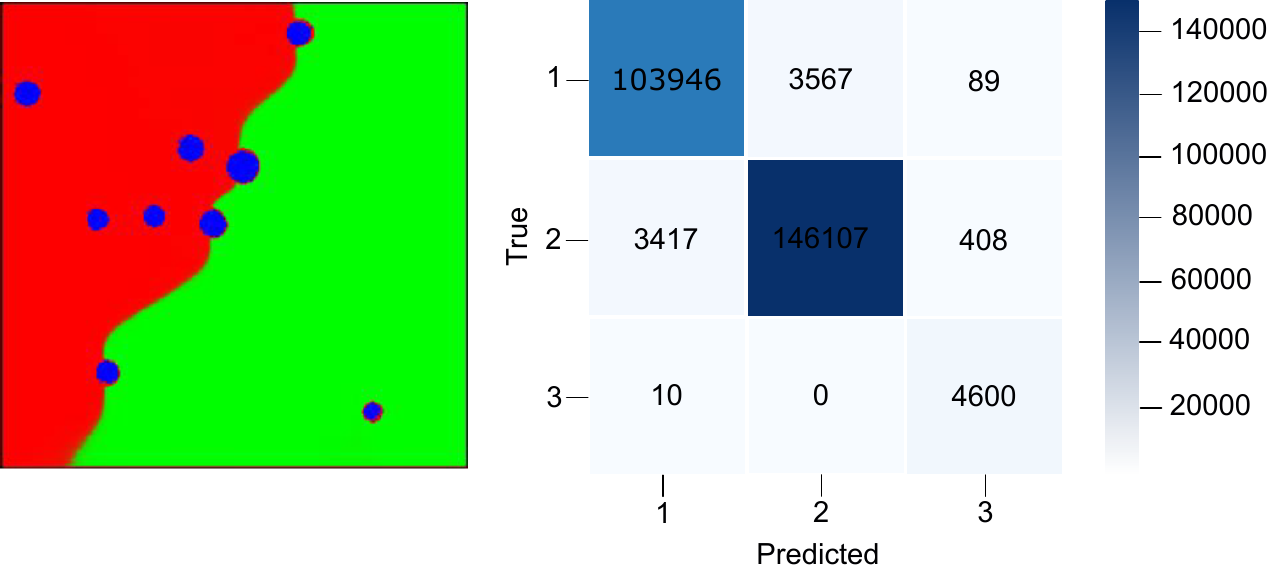}
\caption{Case 2 -- (Left) Predicted crack image for ResNet152 based on electric potential data and (Right) Confusion matrix.}
\label{figure24}
\end{center}
\end{figure}

Similarly, Figures~\ref{figure15} and \ref{figure16} present the image preprocessing pipeline for two representative cases--Case 1 and Case 2--used to generate ground truth crack labels from electric potential simulations. Each figure illustrates three key stages in the data preparation process. The first image (left) shows the original input image, which captures the distribution of holes. The second image (middle) shows the corresponding high-fidelity electric potential map, which reveals spatial voltage gradients induced by the presence of cracks. The final image (right) is the binary mask label, where regions of crack-induced electric field distortion are converted into clear segmentation boundaries.

Figures~\ref{figure17} through \ref{figure24} demonstrate the predictive performance of ResNet-based U-Net architectures for both Case 1 and Case 2 when trained on electric potential data. Each figure consists of two subcomponents: the left panel shows the predicted crack region based on CNN inference, while the right panel presents the corresponding confusion matrix. Specifically, Figures~\ref{figure17} and \ref{figure18} showcase ResNet34 results, Figures~\ref{figure19} and \ref{figure20} correspond to ResNet50, Figures~\ref{figure21} and \ref{figure22} illustrate ResNet101, and Figures~\ref{figure23} and \ref{figure24} present the outputs from ResNet152. These visualizations provide clear comparative insights into the architectural impacts on crack localization accuracy, generalization, and robustness across input conditions.

In Case 1, for example, the predicted crack masks obtained from ResNet34 and ResNet50 models nearly coincide with the ground truth labels, even capturing subtle crack bifurcations and complex curvature paths. The confusion matrices reveal high true positive rates and minimal false negatives, indicating strong segmentation confidence. Importantly, even in regions of gradual voltage gradient decay--typically challenging for threshold-based methods--the models were able to correctly infer crack locations. This demonstrates that electric potentail data encapsulate sufficient physical information about damage evolution, which the CNNs can effectively learn to interpret.

\textcolor{black}{Utilizing electric potential fields as physical signatures during training offers several advantages}. First, their inherently smooth and continuous nature helps convolutional filters extract long-range spatial features without being misled by pixel-level noise. Second, distortions in the voltage gradient field, while subtle, are often more spatially extensive than discrete crack fronts, providing CNNs with a broader contextual field to base decisions upon. Third, because electric potential perturbations result from the disruption of conductive pathways due to cracking, they implicitly encode crack severity and propagation direction. This richness in physical interpretation makes them particularly suited for deep learning-based surrogate modeling.

Despite these advantages, a few limitations are worth noting. False positives can occur near non-cracked defect zones, particularly when local voltage gradients mimic those produced by fractures. Additionally, model confidence can drop in highly congested geometries with overlapping gradients or when defect-induced fields mask actual crack signals. However, through appropriate regularization, ensemble training, and data augmentation, these effects were minimized in our study.

\subsection{Comparative insight}
\label{Comparative}

\textcolor{black}{The preceding sections demonstrated the predictive capabilities of deep convolutional neural networks trained using two distinct physical signatures to guide the mapping from geometry to fracture: phase-field damage variables and electric potential fields.} While both datasets originate from the same underlying finite element simulations of dielectric nanocomposites subjected to mechanical and electrical loading, they represent fundamentally different physical observables. The phase-field field directly encodes the material’s internal damage state as a scalar damage variable, whereas the electric potential field reflects the system’s electrostatic response, perturbed indirectly by the presence and evolution of cracks. This section provides a comparative evaluation of the two approaches, grounded in performance metrics, architectural sensitivity, interpretability, and practical implications for real-world applications.

\subsubsection*{Quantitative metrics comparison}

Tables~\ref{tab:phi_performance} and \ref{tab:v_performance} summarize the training and validation performance of the surrogate models, ranging from ResNet34 to ResNet152, comparing the phase-field-informed and electric potential-informed training strategies, respectively. As evident from these tables, models trained on electric potential data consistently outperformed their phase-field-based counterparts across all ResNet backbones. They not only achieved higher validation F1 and IoU scores but also demonstrated faster and more stable convergence. For instance, the ResNet34 model trained on electric potential data reached a validation IoU of 0.89 within just 50 epochs, whereas the ResNet34 model trained on phase-field data plateaued at an IoU of 0.76 after 75 epochs. This suggests that the electric potential fields provided richer, more learnable features for crack segmentation.

\subsubsection*{Architectural sensitivity}

\textcolor{black}{A key insight from the comparative analysis is the role of network depth in segmentation performance across the two different training strategies.} When trained on electric potential data, deeper ResNet variants--such as ResNet101 and ResNet152--show marginal but consistent improvements in validation accuracy compared to ResNet34. For instance, ResNet152 achieved a validation F1-score of 0.94 and a validation IoU of 0.90, slightly outperforming ResNet34, which achieved 0.93 and 0.89, respectively. These results indicate that deeper architectures can extract and utilize subtle spatial cues embedded in the smooth gradients of electric potential fields.

In contrast, no such improvement is observed with deeper networks when trained on phase-field data. Despite the higher representational capacity of ResNet101 and ResNet152, their performance on phase-field data either stagnates or slightly declines relative to ResNet34. For example, ResNet34 outperformed all other variants with a validation F1-score of 0.82 and an IoU of 0.75, while deeper models failed to surpass this benchmark despite significantly longer training times. This suggests that the highly localized and discontinuous nature of phase-field crack images may not benefit from deeper architectures, which are generally more effective at learning hierarchical or global representations.

The implication is that for phase-field-based segmentation tasks, increasing model depth yields diminishing or even negative returns, potentially due to overfitting to noise or sharp transitions in the data. On the other hand, electric potential data--with their smooth, continuous gradients--support better feature extraction in deeper networks, albeit with marginal gains. 

\subsubsection*{\textcolor{black}{Interpretability and physics-grounded representation}}

While the phase-field approach is directly rooted in the mechanics of fracture, its effectiveness in training CNNs is often constrained by the highly localized and discontinuous nature of the damage variable. In contrast, the electric potential distribution represents a continuous, global field influenced by both material fracture and the underlying structural geometry. This yields significant advantages for deep learning; specifically, electric potential fields offer context-rich features by providing diffuse gradients that inherently reflect the presence of cracks, enabling the network to capture complex spatial dependencies more effectively.

\textcolor{black}{The CNN's ability to segment cracks from these potential maps is physically grounded in the "insulating crack" boundary condition within the electromechanical framework. Regions where the damage variable $\mathfrak{d} \approx 1$ behave as low-permittivity channels. Physically, while the electric field penetrates the crack volume--modeled here as a medium with low but non-zero permittivity--the sharp contrast with the dielectric bulk matrix generates distinct field perturbations. Consequently, the electric potential and electric field are forced to bend sharply around the crack, creating high-gradient "corridors" and distorted equipotential contours that serve as unique physical signatures of the crack path.} 

\textcolor{black}{Consistent with the findings in \cite{Ref050}, the local field distribution is highly sensitive to the electrical boundary conditions at the crack faces. While the present study adopts a strong permittivity degradation with fixed electrode potentials, different physical regimes--such as short-circuit or partially conducting cracks--would yield distinct field signatures. Because these electro-mechanical distortions extend beyond the immediate crack tip, the training data provide greater spatial coverage and a more balanced class distribution, leading to enhanced numerical stability and faster convergence.}

\textcolor{black}{Ultimately, the network does not merely perform pattern recognition; it leverages these physically consistent field perturbations to reconstruct the high-fidelity ground-truth crack paths obtained from the phase-field simulations. The ability of the deep learning framework to generalize across varying electrical boundary conditions or to enforce perfectly insulating cracks remains a compelling direction for future work.}

\subsubsection*{Limitations and challenges}

Despite its superior performance, electric potential data-based learning is not without limitations. One challenge lies in interpretability: while phase-field values directly correlate with damage, interpreting voltage-induced field distortions as "cracks" requires physical insight and a well-trained model. Furthermore, in complex geometries or low signal-to-noise regimes, voltage gradients may be too diffuse to pinpoint precise crack boundaries, potentially affecting precision.

\textcolor{black}{A significant physical limitation arises in multi-crack scenarios due to the "shielding effect" inherent in global elliptic fields like the electric potential. As observed in Figure~\ref{figure2}, while the framework accurately captures dominant fracture paths, it may overlook secondary satellite cracks or non-propagating branches. Because the electric potential is a global field, its distribution is primarily dictated by the most significant geometric discontinuities. Dominant cracks can physically mask the signatures of smaller branches, leaving them with insufficient field gradients for the network to reliably segment. This suggests that while the current modality is highly sensitive to overall structural connectivity, it may require augmentation--such as multi-modal sensing incorporating local strain fields--to resolve fine-scale secondary cracking events.}

Models trained on phase-field data, although lower in raw accuracy, benefit from clearer interpretability and direct fracture encoding, which may be valuable in verification or certification workflows where transparency is crucial.

\subsubsection*{Implications for real-world deployment}
From a practical standpoint, the comparative analysis reveals that models trained on electric potential data are more amenable to deployment in real-time structural health monitoring (SHM) systems, especially in dielectric or piezoelectric materials. Voltage fields are easier to measure using embedded sensors or surface electrodes, whereas obtaining high-resolution fracture fields requires full-field simulation or experimental reconstruction, both of which are computationally intensive or infeasible in situ.

Thus, a key implication of this study is that electric potential fields can serve as a physically grounded, sensor-accessible proxy for internal fracture, and when combined with neural networks, offer a powerful hybrid modeling framework for rapid and accurate crack detection.

\subsection{\textcolor{black}{Computational efficiency and speed-up analysis}}
\label{Efficiency}

\textcolor{black}{A primary motivation for developing the hybrid deep learning framework is to overcome the prohibitive computational cost associated with high-fidelity, fully coupled electromechanical phase-field simulations. To evaluate the efficiency of the proposed surrogate model, we conducted a direct comparison of the execution times for a single crack propagation case.}

\textcolor{black}{The reference high-fidelity simulations, performed using the finite element method on a workstation equipped with 32 CPU cores, required approximately 15 seconds per case to converge. In contrast, once trained, the ResNet-U-Net model completes a single inference (prediction) in approximately 0.1 seconds on an NVIDIA RTX 3090 GPU. This represents a significant 150-fold speed-up in prediction time.} 

\textcolor{black}{Critically, this reduction in computational burden is achieved while maintaining high spatial accuracy, as evidenced by an IoU of approximately 0.9 for the electric potential-based predictions. Such a dramatic increase in efficiency demonstrates the potential of the proposed framework for real-time applications, such as structural health monitoring and iterative design optimization.}

\section{Conclusion and future work}
\label{Conclusion}

This study introduced a hybrid computational framework that integrates electromechanical phase-field fracture modeling with deep learning–based surrogate models to predict crack propagation in boehmite-reinforced dielectric nanocomposite plates. The methodology is rooted in \textcolor{black}{high-fidelity physics-based simulations} that account for both stiffness degradation and permittivity loss due to fracture. Two physically meaningful observables--the phase-field damage variable and the electric potential distribution--were used as training data for ResNet-U-Net convolutional neural networks, enabling pixel-wise crack segmentation.

The results reveal several key findings:

\begin{itemize}
\item CNN models trained on electric potential fields consistently outperformed those trained on phase-field fields across all evaluated architectures, including ResNet34, ResNet50, ResNet101, and ResNet152.
\item Electric potential data yielded higher F1 and IoU scores, faster convergence, and greater robustness, owing to their smooth, continuous nature and implicit encoding of crack-induced electrostatic distortions.
\item Phase-field data, while physically explicit in encoding fracture, posed challenges related to class imbalance, sharp discontinuities, and training instability.
\item Shallower networks such as ResNet34 achieved performance comparable to deeper variants when trained on electric potential data, making them favorable for real-time and resource-constrained applications.
\item Visual analysis and confusion matrices confirmed the models' ability to accurately detect complex crack topologies, including bifurcations and micro-cracks, especially in electric potential–based training.
\item \textcolor{black}{The proposed framework achieves a significant reduction in computational cost; while a high-fidelity electromechanical phase-field simulation requires approximately 15 seconds per case on 32 CPU cores, the trained CNN completes the prediction in 0.1~s on a GPU, representing a 150-fold speed-up.}
\end{itemize}

The comparative analysis underscores that electric potential fields can serve as reliable and sensor-accessible proxies for internal fracture states, opening pathways for machine learning--enabled real-time structural health monitoring. This may provide an efficient alternative to computationally expensive full-field simulations, without compromising on prediction fidelity.

Future research directions include:

\begin{itemize}
\item Integration of recurrent or attention-based deep learning architectures (e.g., ConvLSTM, Transformers) to capture temporal evolution of crack growth in time-dependent loading scenarios.
\item Extension of the framework to 3D simulations and volumetric segmentation, allowing real-world deployment in thick-walled components and layered structures.
\item Experimental validation using digital image correlation (DIC) and embedded voltage sensors to benchmark model predictions against physical measurements \textcolor{black}{and assess reliability under real-world loading conditions beyond simulation-based training}.
\item Incorporation of physics-informed neural networks (PINNs) to embed governing partial differential equations (PDEs) directly into the learning process, further improving generalization with limited data.
\item \textcolor{black}{Exploration of transfer learning across diverse geometry classes (e.g., notches and varying aspect ratios) and loading conditions, enabling the development of universal fracture prediction models for smart materials that generalize beyond the specific patterns used in this study.}
\item \textcolor{black}{Invesitigation of the impact of alternative electrical boundary conditions and crack-face properties--such as perfectly insulating or partially conducting cracks--as discussed in recent multiferroic phase-field studies \cite{Ref050}. Additionally, for more complex multi-crack topologies, we intend to implement granular spatial-residual error maps to further investigate the local architectural limitations of the surrogate model.}
\end{itemize}

In conclusion, this study demonstrates the potential of combining physical modeling and data-driven learning to advance intelligent diagnostics in dielectric materials. The hybrid methodology bridges simulation and sensing, offering scalable, accurate, and interpretable tools for next-generation smart structures.

\section*{Acknowledgment}
This work stems from the research project “Functionalized, Multi-Physically Optimized Adhesive Systems for Inherent Structural Monitoring of Rotor Blades” (Func2Ad – Funktionalisierte, multiphysikalisch optimierte Klebstoffsysteme für die inhärente Strukturüberwachung von Rotorblättern), funded by the Federal Ministry for Economic Affairs and Climate Action, Germany (Grant No. 03EE3069A). The authors gratefully acknowledge this financial support. They also acknowledge the use of the LUIS scientific computing cluster, Germany, funded by Leibniz Universität Hannover, the Lower Saxony Ministry of Science and Culture (MWK), and the German Research Foundation (DFG).

\section*{Compliance with ethical standards}

\noindent \textbf{Data availability statement:}\\
The data supporting the findings of this study are available from the corresponding author upon reasonable request.

\noindent \textbf{Funding statement:}\\
This research did not receive any specific grant from funding agencies in the public, commercial, or not-for-profit sectors.

\noindent \textbf{Conflict of interest disclosure:}\\
The authors declare that they have no known competing financial interests or personal relationships that could have appeared to influence the work reported in this paper.

\noindent \textbf{Ethics approval statement:}\\
Not applicable.

\noindent \textbf{Patient consent statement:}\\
Not applicable.

\noindent \textbf{Permission to reproduce material from other sources:}\\
Not applicable.

\noindent \textbf{Clinical trial registration:}\\
Not applicable.

\noindent \textbf{Declaration of AI-assisted technologies in the writing process:}\\
During the preparation of this manuscript, the authors utilized AI-based tools to assist with language refinement and grammar checking. Following the use of these tools, the authors thoroughly reviewed and edited the content, and take full responsibility for the final version of the publication.

\bibliographystyle{elsarticle-num}

\bibliography{references}

\end{document}